\documentclass[aps,prd,twocolumn,superscriptaddress,nofootinbib,floatfix]{revtex4-2}
\usepackage[T1]{fontenc}
\usepackage{lmodern}
\usepackage{amsmath}
\usepackage{amssymb}
\usepackage{amsfonts}
\usepackage{bm}
\usepackage{slashed}
\usepackage{booktabs}
\usepackage{graphicx}
\usepackage{placeins}
\usepackage[colorlinks=true,allcolors=blue]{hyperref}

\begin{document}

\title{Matter Swapping between Two Branes:
Conditions for a Local Type IIB D3--D7--D3 Mechanism of Charged-Fermion Mixing via E3 Instantons}
\author{Micha\"el Sarrazin}
\email{michael.sarrazin@ac-besancon.fr}
\affiliation{Universit\'e Marie et Louis Pasteur, CNRS, Institut UTINAM (UMR 6213),
\'Equipe de Physique Th\'eorique, F-25000 Besan\c{c}on, France}

\begin{abstract}
Effective infrared two-sheeted descriptions of fermion mixing between
braneworlds have emerged in several distinct realizations and lead to
testable matter-swapping effects, motivating low-energy searches and
cosmological applications. The recurrence of the same low-energy fermion
operator motivates investigating whether this structure can also arise
from a D-brane mechanism. We investigate the conditions under which a
local Type IIB D3$_+$--D7--D3$_-$/E3 configuration can generate
perturbatively forbidden interbrane mixing. Assuming the required
charged zero-mode spectrum, neutral instanton measure and non-vanishing
charged disk amplitudes, we derive an E3-induced chiral mass term with
scalar and pseudoscalar components. The latter reproduces the
characteristic mixing structure of the two-sheeted fermion operator,
while D7 Wilson transport supplies the gauge-invariant relative
connection between the two brane sectors. We establish the corresponding
gauge matching for transitions between stationary states at resonance.
This yields a conditional matching to the two-sheeted operator within
the effective field theory. A separate orbifold example checks the charged
zero-mode representations and the compensating instanton charge.
Whether all the required instanton ingredients coexist with two
independent massive D3--D7 matter sectors at separated endpoints in a
single local configuration remains open. We outline a model-building
roadmap toward such a configuration and its compact phenomenological
embedding. Such an embedding requires electromagnetic sequestering of
the two fermion sectors and a sufficiently light non-uniform D7 gauge
mode mediating a long-range relative interaction. In such a
compactification, the mixing parameter $g$ would be determined by
microscopic instanton data and matter-field normalization, providing a
connection to low-energy constraints.
\end{abstract}

\maketitle

\section{Introduction}

\label{sec:intro}

In a Universe containing two parallel branes --- one corresponding to our
visible universe, the other to a hidden universe --- matter fields localized
on different branes can mix. This possibility is not merely formal: it leads
to experimentally testable signatures that are already under active
investigation, including neutron disappearance and regeneration in reactor
and ultracold-neutron experiments \cite%
{PLB2012,PRD2015,PLB2016,MURMUR,STEREO}, invisible decays of
positronium and quarkonia \cite{IJMPA2006,IJMPA2020b}, and a novel
baryogenesis mechanism \cite{PRD2024,EPJC2025}. The low-energy dynamics of
such a system can be efficiently described by a Dirac field residing on a
noncommutative two-sheeted spacetime $M_{4}\times Z_{2}$ --- a property we
hereafter refer to as the Two-Brane--Noncommutative Two-Sheeted Spacetime
(TBTS) Correspondence \cite{PLB2005,ActaPolon2005,IJMPA2007,PRD2010,IJMPA2019,IJMPA2020a}.

String theory realizes braneworld scenarios through D$p$-branes, $p$%
-dimensional surfaces on which open strings can end \cite%
{Polchinski:1998rq,Bachas:1998rg,Polchinski:1995mt}. Closely related ideas
appear in the localization of Standard Model fields on topological defects 
\cite%
{RubakovShaposhnikov,Akama1983,Pavsic1986,Visser1985,DubovskyRubakovTinyakov2000,ArkaniHamedSchmaltz2000,RingevalPeterUzan2002,DGS}
and in warped geometries \cite{RandallSundrum1,RandallSundrum2}. In these
contexts, two-brane configurations have also been invoked in physically
motivated settings ranging from early-Universe cosmology and Big-Bang
scenarios \cite%
{KhouryOvrutSteinhardtTurok2001,GibbonsLuPope2005,BattefeldPeter2015} to
interpretations of dark matter or dark energy \cite%
{ArkaniHamedDimopoulosKaloperDvali2000,BraxVandeBruckDavis2004,KoivistoWillsZavala2014}%
. They are therefore not introduced here simply as a formal doubling of
spacetime, but as a class of braneworld configurations with independent
cosmological and phenomenological motivations.

Because a complete ultraviolet description of Standard Model dynamics in a braneworld
setting is often prohibitively complex, the underlying two-brane/two-sheeted
correspondence was formulated as an infrared equivalence of a two-brane
Universe \cite{PLB2005,ActaPolon2005,IJMPA2007}. It has since been established 
in a domain-wall model \cite{PRD2010}, in phenomenological bulk-fermion models \cite{IJMPA2019,IJMPA2020a}
and has a solid-state analog \cite{EPJB2014}. A recurrent
feature of these realizations is the same infrared fermion-operator
structure: details such as the number of bulk fields, dimensionality, or
warping are absorbed into a geometrical coupling constant $g$ between the
visible and hidden sectors $\psi_+$ and $\psi_-$. Since this two-sheeted
description already gives rise to experimentally testable predictions,
embedding its underlying mechanism within string theory offers an unusual
opportunity to connect an underlying string braneworld construction with
laboratory phenomenology.

Here we ask whether the charged-fermion mixing encoded by this infrared
two-sheeted description can arise from a D-brane mechanism. Our aim is to
identify the instanton-sector conditions under which a local Type IIB
D3--D7--D3/E3 mechanism generates the required mixing operator and to
establish its gauge matching. With the specified instanton measure and
non-vanishing charged disk couplings, Grassmann saturation yields an
off-diagonal chiral mass whose Dirac decomposition contains the
characteristic pseudoscalar mixing and its scalar companion. D7 Wilson
transport supplies the relative gauge connection between the fermion
endpoint loci, allowing the gauge matching to be established for stationary
resonant matrix elements.

The scope of this work is the local instanton mechanism and its matching 
to the two-sheeted fermion operator under explicitly stated microscopic 
assumptions. Constructing an orientifold background that realizes these assumptions 
together with the massive, separated-endpoint matter configuration is a distinct 
model-building task, as is its subsequent embedding into a globally consistent 
compactification. These tasks lie beyond the scope of the present 
analysis. In particular, we do not demonstrate that the required chiral
 E3--D3 zero-mode spectrum, compensating instanton charge, neutral 
instanton measure and two independent massive D3--D7 sectors can be 
realized simultaneously in a single consistent local configuration, nor 
do we evaluate the complete charged disk amplitudes. The results should 
therefore be read as a conditional operator matching rather than as an explicit 
string realization. Section~\ref{sec:roadmap} summarizes the corresponding realization 
conditions and the additional requirements for phenomenological applications, providing a model-building roadmap for future work.

To make the target of the string construction explicit, we briefly recall
the effective Lagrangian governing fermion dynamics on the noncommutative
two-sheeted spacetime \cite{PLB2005,IJMPA2007,PRD2010,IJMPA2019,IJMPA2020a}. Writing
the matter field as a two-sheeted doublet $\bm{\Psi}=(\psi _{+},\psi _{-})^{T}$,
one has 
\begin{equation}
\mathcal{L}_{M_{4}\times Z_{2}}\sim \overline{\bm{\Psi}}\left( i{\slashed{D}}%
_{A}-M\right) \bm{\Psi},  \label{eq:L2}
\end{equation}%
with the Dirac operator 
\begin{align}
iD\!\!\!\!\slash\,_{A}-M& =%
\begin{pmatrix}
\mathcal{D}_{+}-m & ig\gamma ^{5}-im_{r} \\[3pt] 
ig\gamma ^{5}+im_{r} & \mathcal{D}_{-}-m%
\end{pmatrix}%
,  \notag \\
\text{with }\mathcal{D}_{\pm }& \equiv i\gamma ^{\mu }(\partial _{\mu
}+iqA_{\mu }^{\pm }).  \label{eq:Dop}
\end{align}%
Here $A_{\mu }^{\pm }$ are the electromagnetic four-potentials on each
brane, $m$ is the brane-localized mass, $g$ the pseudoscalar geometrical
mixing, and $m_{r}$ an antisymmetric scalar entry. We take $g,m_r\in\mathbb{R}$ throughout. The derivative operators
acting on $M_{4}$ and $Z_{2}$ are $D_{\mu }=\mathbf{1}_{8\times 8}\partial
_{\mu }$ ($\mu =0,1,2,3$) and $D_{5}=ig\sigma _{2}\otimes \mathbf{1}%
_{4\times 4}$, respectively, and the Dirac operator acting on $M_{4}\times
Z_{2}$ is defined as ${\slashed{D}}=\Gamma ^{N}D_{N}=\Gamma ^{\mu }D_{\mu
}+\Gamma ^{5}D_{5}$, where $\Gamma ^{\mu }=\mathbf{1}_{2\times 2}\otimes
\gamma ^{\mu }$ and $\Gamma ^{5}=\sigma _{3}\otimes \gamma ^{5}$. The
matrices $\gamma ^{\mu }$ and $\gamma ^{5}=i\gamma ^{0}\gamma ^{1}\gamma
^{2}\gamma ^{3}$ are the usual Dirac matrices and $\sigma _{k}$ ($k=1,2,3$)
are the Pauli matrices. This is the standard Dirac operator of
noncommutative $M_{4}\times Z_{2}$ geometry \cite%
{ConnesLott1991,Connes1994,ChamseddineConnes1996,LizziEtAl1997,LizziEtAl1998,GraciaBondiaIochumSchucker1998,KaseMoritaOkumura1999,KaseMoritaOkumura2001,MacesanuWali2006,VietWali1996,VietWali2003}%
. Its electromagnetic gauge group is $U(1)_{+}\times U(1)_{-}$, so that in the unbroken
theory the two photons are localized independently \cite{PRD2010}. The electromagnetic field $%
\slashed{A}\sim diag(\gamma ^{\mu }A_{\mu }^{+},\gamma ^{\mu }A_{\mu
}^{-})$ is then introduced in the Dirac equation through ${\slashed{D}}%
_{A}\rightarrow {\slashed{D}}+iq\slashed{A}$ \cite{PRD2010}. In the nonrelativistic limit, the pseudoscalar mixing gives rise to
the spin-dependent, magnetic-vector-potential-dependent contribution to the
interbrane Hamiltonian 
\begin{equation}
H_{\mathrm{cm}}=%
\begin{pmatrix}
0 & -ig\,\hat{\bm\mu }\cdot (\mathbf{A}_{+}-\mathbf{A}_{-}) \\[2pt] 
ig\,\hat{\bm\mu }\cdot (\mathbf{A}_{+}-\mathbf{A}_{-}) & 0%
\end{pmatrix}%
,  \label{eq:Hcm-intro}
\end{equation}%
which drives the swapping between branes \cite{PLB2012,PRD2015,PLB2016,MURMUR,STEREO,PRD2010,EPJC2012} and 
where $\mathbf{A}_{\pm}$ are the magnetic vector potentials in each brane, 
and $\hat{\bm\mu }$ is the magnetic moment operator of the fermion.
We emphasize this spin-dependent relative-potential structure because it
is characteristic of the two-sheeted description and constitutes the
matching target of this work. Scalar mixing alone does not identify this
structure, although its contribution must be retained when evaluating
physical transitions.
The dependence of Eq.~(\ref{eq:Hcm-intro}) on the relative potential is not
an artifact of the nonrelativistic reduction: as shown in Sec.~\ref%
{sec:fieldtheory}, it follows from an exact identity satisfied by the mixed
axial current of the two-sheeted doublet in the stationary resonant matrix
element. The scalar contribution associated with $m_r$ is retained throughout. Establishing 
whether magnetic swapping dominates over this contribution requires evaluating 
the transition matrix elements for the states and gauge background considered (see Sec.~\ref{sec:fieldtheory}).

Since Eqs.~(\ref{eq:L2})--(\ref{eq:Hcm-intro}) describe charged matter
localized in two distinct brane sectors, we will consider the minimal local
matter content needed for this question: one D3$_{+}$ brane, one D3$_{-}$
brane, and one common D7 gauge brane. In a D-brane description, gauge charges
are carried by Chan--Paton degrees of freedom at open-string endpoints.
Throughout, N and D denote Neumann and Dirichlet boundary conditions,
respectively and NN and DD denote Neumann--Neumann and Dirichlet--Dirichlet
conditions, while ND and DN denote the two orientations of mixed
Neumann--Dirichlet conditions. When only the number of mixed directions
matters, we use ``ND'' generically, irrespective of orientation. We
therefore realize each fermion in a $37+73$ open-string sector: the D3
endpoint distinguishes the visible and hidden sectors (i.e. $\psi _{+}$ and $%
\psi _{-}$ respectively), whereas the common D7\footnote{%
A D7 is the natural choice here because a D3--D7 pair has four ND directions
and provides the standard supersymmetric $37+73$ matter sector, whereas
D3--D5 and D3--D9 pairs have respectively two and six ND directions and do
not furnish the same supersymmetric sector \cite%
{Polchinski:1998rq,Bachas:1998rg}. A single D7 can moreover couple both
D3--D7 sectors to the same gauge field.} makes both Dirac fermions couple to
the same Abelian gauge field. Moreover, the string construction in the
following is not intended to reproduce the product gauge group $%
U(1)_{+}\times U(1)_{-}$. Instead, $A_{\mu }^{\pm }$ will arise as the
values of a common D7 Abelian connection at the two fermion endpoint loci.
The matching performed here therefore concerns the relative gauge connection
entering the transition operator, rather than a complete realization of the
two-photon gauge sector \footnote{%
The possibility of obtaining a long-range relative interaction depends on
the D7 compactification spectrum. This issue is discussed in Sec.~\ref%
{sec:modes}.}.

This structure makes sector mixing non-trivial. Indeed, in this
minimal setup, changing the D3 endpoint changes the Chan--Paton sector, so
that the direct cross-brane bilinear is perturbatively forbidden. A virtual
exchange of the massive strings stretched between the two D3s does not by
itself remove this obstruction: without additional charged insertions or a
vacuum expectation value, perturbative amplitudes remain subject to the same
Chan--Paton selection rule. This motivates a non-perturbative E3-instanton
channel, in which suitable charged zero modes can provide
Chan--Paton-allowed couplings whose saturation generates the otherwise
forbidden off-diagonal mass term while preserving exact gauge invariance.

To isolate this microscopic mechanism, we deliberately restrict the analysis
to one elementary charged Dirac fermion in each sector and introduce neither
a color group nor hadronic matching.
The D3 trace $U(1)$ factors are not identified with electromagnetism. They
are nevertheless not spectators of the mechanism: the relative combination
$U(1)_{3+}-U(1)_{3-}$ carries the perturbative selection rule, and its
compensation by the E3 action is a \emph{local} condition on the
instanton--D3 sector, discussed in Secs.~\ref{sec:CPobstruction}
and~\ref{sec:chargedmodes}.

The paper proceeds as follows. Section~\ref{sec:fieldtheory} derives, within
the TBTS framework, the matter-swapping transition amplitude that the string
construction must reproduce. Section~\ref{sec:string} introduces the minimal
D3$_+$--D7--D3$_-$ geometry, the equal physical but opposite holomorphic D3--D7
masses, and the perturbative Chan--Paton obstruction. Section~\ref%
{sec:instanton} develops the local E3 mechanism, shows how the otherwise
forbidden off-diagonal mixing is generated from the instanton and charged-disk
data, and states its realization conditions, including the charged zero-mode
structure and the finite-separation world-sheet contribution. Section~\ref{sec:gamma5}
converts the resulting string mass matrix to the positive-mass Dirac basis
and shows how the characteristic TBTS $i\gamma^5$ structure emerges. Section~%
\ref{sec:relative} then Wilson-completes this string-generated operator and
performs its leading soft matching to the field-theory target of
Sec.~\ref{sec:fieldtheory}, while discussing the conditions under which the
relative interaction can remain long-ranged. Finally, before the conclusion, Section~\ref{sec:roadmap} 
outlines a model-building roadmap for future work, summarizing the conditions 
required to realize the local mechanism and the further steps toward a 
globally consistent compactification with low-energy phenomenological applications.

\section{Interbrane transition amplitude and the origin of the relative
potential}

\label{sec:fieldtheory}

Before turning to the string construction we establish, within the classical
field theory defined by Eqs.~(\ref{eq:L2}) and~(\ref{eq:Dop}), the
mixed-current identity relevant at resonance and the associated leading Born
matrix element in an external gauge background, i.e. the matter swapping
amplitude. The point is to make explicit how the relative potential $\mathbf{%
A}_{+}-\mathbf{A}_{-}$ emerges although the off-diagonal block of (\ref%
{eq:Dop}) contains no gauge field, to fix the relativistic target and
normalization used in the string matching, and to show why this structure is
specific to the pseudoscalar mixing.

Throughout we use the metric $(+,-,-,-)$, $\gamma ^{5}=i\gamma ^{0}\gamma
^{1}\gamma ^{2}\gamma ^{3}$, $\beta =\gamma ^{0}$, $\alpha ^{i}=\gamma
^{0}\gamma ^{i}$ and $\bm{\Sigma }=\gamma ^{5}\bm{\alpha }$. We write 
\begin{equation}
D_{\mu }^{\pm }=\partial _{\mu }+iqA_{\mu }^{\pm },\qquad \mathcal A_{\mu
}\equiv A_{\mu }^{+}-A_{\mu }^{-},  \label{eq:relpot}
\end{equation}
$\mathcal A_{\mu }$ being the connection of the antidiagonal (relative) $%
U(1)$ of $U(1)_{+}\times U(1)_{-}$. In this two-sheeted description it is a
relative gauge connection rather than a gauge-invariant quantity by itself.
The string construction of Sec.~\ref{sec:relative} will provide the
Wilson-completed D7 representation of this same physical relative connection.

\subsection{First-order amplitude}

Expanding (\ref{eq:L2}) with (\ref{eq:Dop}) gives 
\begin{align}
\mathcal{L}=& \sum_{s=\pm }\overline{\psi }_{s}\left( i\gamma ^{\mu }D_{\mu
}^{s}-m\right) \psi _{s}  \notag \\
& +\overline{\psi }_{-}\left( ig\gamma ^{5}+im_{r}\right) \psi _{+}+\mathrm{%
\ h.c.},  \label{eq:Lexpanded}
\end{align}%
The displayed off-diagonal term and its Hermitian conjugate together form a Hermitian interaction. In the Hamiltonian form, we
get $i\partial _{t}\bm{\Psi}=H\bm{\Psi} $ such that%
\begin{equation}
H=%
\begin{pmatrix}
H_{+} & \mathcal{W}^{\dagger } \\[2pt] 
\mathcal{W} & H_{-}%
\end{pmatrix}%
,\qquad H_{\pm }=\bm{\alpha }\cdot \bm{\pi }_{\pm }+\beta m+qA_{\pm }^{0},
\label{eq:Hfull}
\end{equation}%
with $\bm{\pi }_{\pm }=\mathbf{p}-q\mathbf{A}_{\pm }$ and where 
\begin{equation}
\mathcal{W}=-ig\,\beta \gamma ^{5}-im_{r}\,\beta  \label{eq:W}
\end{equation}%
is the generator of the $+\rightarrow -$ transition. Let $\psi _{i}$ and $%
\psi _{f}$ be exact stationary solutions of $H_{+}$ and $H_{-}$ with
energies $E_{i}$ and $E_{f}$, i.e.\ solutions of 
\begin{equation}
\left( i\gamma ^{\mu }D_{\mu }^{+}-m\right) \psi _{i}=0,\qquad \left(
i\gamma ^{\mu }D_{\mu }^{-}-m\right) \psi _{f}=0.  \label{eq:eom0}
\end{equation}%
To first order in the mixing the transition amplitude is $\mathcal{T}%
_{fi}=-i\int \!dt\,\langle f|\mathcal{W}|i\rangle e^{i(E_{f}-E_{i})t}$ with 
\begin{equation}
\langle f|\mathcal{W}|i\rangle =-ig\!\int \!d^{3}x\,\overline{\psi }%
_{f}\gamma ^{5}\psi _{i}-im_{r}\!\int \!d^{3}x\,\overline{\psi }_{f}\psi
_{i}.  \label{eq:bare}
\end{equation}%
Equation~(\ref{eq:bare}) contains no gauge potential. The potentials are
nevertheless hidden in it, because $\psi _{i}$ and $\psi _{f}$ obey
different equations in (\ref{eq:eom0}). Making this dependence explicit
requires an identity relating the pseudoscalar transition bilinear to the
two gauge connections.

\subsection{Mixed axial identity}

Because the transition matrix element in Eq.~(\ref{eq:bare}) contains the
pseudoscalar bilinear $\overline{\psi}_f\gamma^5\psi_i$, the natural
quantity to examine is the mixed axial current built from the initial and
final states. Its divergence can be evaluated using their respective Dirac
equations and thereby exposes how the two gauge connections enter the
transition amplitude. We therefore define 
\begin{equation}
J_5^\mu \equiv \overline{\psi}_f\gamma^\mu\gamma^5\psi_i.  \label{eq:J5}
\end{equation}

Using (\ref{eq:eom0}) in the form $\gamma ^{\mu }\partial _{\mu }\psi
_{i}=-i(m+q\slashed{A}_{+})\psi _{i}$ and $\partial _{\mu }\overline{\psi }%
_{f}\gamma ^{\mu }=i\overline{\psi }_{f}(m+q\slashed{A}_{-})$, and
anticommuting $\gamma ^{5}$ through $\gamma ^{\mu }$ twice, one finds from (%
\ref{eq:J5}) 
\begin{equation}
\partial _{\mu }J_{5}^{\mu }=2im\,\overline{\psi }_{f}\gamma ^{5}\psi
_{i}+iq\,\overline{\psi }_{f}\gamma ^{5}\slashed{\mathcal A}\,\psi _{i}.
\label{eq:axialid}
\end{equation}%
This identity --- the two-brane analog of the partially conserved
axial-current relation --- is the crux of the argument. The two
anticommutations act on the two branes with opposite outcomes, so that $%
A_{+} $ and $A_{-}$ enter with a relative sign: only the difference $%
\mathcal A_{\mu }$ survives, and it does so as a consequence of the
pseudoscalar nature of the mixing alone.

From Eq. (\ref{eq:axialid}), the pseudoscalar bilinear becomes 
\begin{equation}
\overline{\psi }_{f}\gamma ^{5}\psi _{i}=-\frac{q}{2m}\overline{\psi }%
_{f}\gamma ^{5}\slashed{\mathcal A}\,\psi _{i}+\frac{1}{2im}\partial _{\mu
}J_{5}^{\mu },  \label{eq:solved}
\end{equation}%
where for normalizable states the spatial part of $\int d^{3}x\,\partial
_{\mu }J_{5}^{\mu }$ is a surface term and drops, while the remainder is $%
\partial _{t}\!\int \!d^{3}x\,\psi _{f}^{\dagger }\gamma ^{5}\psi
_{i}\propto i(E_{f}-E_{i})$, which vanishes at resonance, i.e. when $%
E_{i}=E_{f}.$ The resonance condition refers to the complete stationary
eigenenergies of $H_+$ and $H_-$. In particular, both the electrostatic
potentials $A_\pm^0$ and the spatial gauge backgrounds entering 
through $\bm{\pi }_{\pm }=\mathbf{p}-q\mathbf{A}_{\pm }$ can contribute to the detuning.
Resonance therefore does not require $A_\mu^+=A_\mu^-$: a non-zero relative
gauge potential can be compatible with $E_i=E_f$. For the purpose of the
present derivation, we assume that this complete-energy resonance condition
is satisfied. At resonance, the spatially integrated divergence vanishes
exactly, and Eq.~(\ref{eq:solved}) becomes an exact relation for the
corresponding stationary matrix element.

\subsection{Covariant amplitude and its nonrelativistic limit}

Inserting now Eq.~(\ref{eq:solved}) into Eq.~(\ref{eq:bare}) gives 
\begin{align}
\langle f|\mathcal{W}|i\rangle ={}& \frac{iqg}{2m}\!\int \!d^{3}x\,\bar{\psi}%
_{f}\gamma ^{5}\slashed{\mathcal A}\psi _{i}-\frac{g}{2m}\!\int
\!d^{3}x\,\partial _{\mu }J_{5}^{\mu }  \notag \\
& -im_{r}\!\int \!d^{3}x\,\bar{\psi}_{f}\psi _{i}.  \label{eq:masteroffres}
\end{align}%
At resonance, this reduces to 
\begin{equation}
\langle f|\mathcal{W}|i\rangle =\frac{iqg}{2m}\!\int \!d^{3}x\,\overline{%
\psi }_{f}\gamma ^{5}\slashed{\mathcal A}\,\psi _{i}-im_{r}\!\int \!d^{3}x\,%
\overline{\psi }_{f}\psi _{i}.  \label{eq:master}
\end{equation}%

In a plane-wave representation,
$\psi_i=u_+(p,r)e^{-ip\cdot x}$ on brane $+$ and
$\psi_f=u_-(p',r')e^{-ip'\cdot x}$ on brane $-$,
the Fourier-space kernel of the effective resonant transition
operator reads
\begin{align}
\mathcal{M}_{+\rightarrow -}={}&
\frac{iqg}{2m}\,
\overline{u}_{-}(p',r')\gamma^5
\slashed{\mathcal A}(k)u_{+}(p,r)
\notag\\
&-im_r\,\widetilde{1}(k)\,
\overline{u}_{-}(p',r')u_{+}(p,r),
\label{eq:spinoramp}
\end{align}
where $k=p'-p$, $\mathcal A_\mu(k)$ is the Fourier transform
of the relative potential, and
$\widetilde{1}(k)=(2\pi)^4\delta^{(4)}(k)$ is the transform
of the constant scalar coefficient.
For a stationary background, the common energy-conserving
factor can be extracted, leaving a spatial momentum delta
function in the scalar term. The analogous expression holds
for $v$ spinors.

Equation~(\ref{eq:master}) is the exact stationary matrix
element at resonance to first order in the interbrane mixing.
Equation~(\ref{eq:spinoramp}) displays the corresponding
effective kernel evaluated between plane waves.
Its pseudoscalar term gives the leading contribution linear
in the relative gauge background. The scalar kernel has no
explicit background dependence, but its matrix element
between exact states generally does: corrections from the
background-induced deformation of these states are retained
in Eq.~(\ref{eq:master}) and are not included in the
plane-wave kernel above.
The pseudoscalar term is the relativistic spinor-level
counterpart of Eq.~(\ref{eq:Hcm-intro}).
To make this connection explicit, let us separate the temporal
and spatial parts of $\slashed{\mathcal A}$,
\begin{equation}
\overline{\psi }_{f}\gamma ^{5}\slashed{\mathcal A}\psi _{i}=\psi
_{f}^{\dagger }\left[ \bm{\Sigma }\cdot \bm{\mathcal A}-\gamma ^{5}\mathcal%
A^{0}\right] \psi _{i},  \label{eq:decomp}
\end{equation}%
so that the interbrane block of the Hamiltonian in Eq.~(\ref{eq:Hfull}) may
be replaced by the effective operator 
\begin{equation}
\mathcal{W}_{\mathrm{eff}}=ig\left[ \hat{\bm\mu }_{\Sigma }\cdot %
\bm{\mathcal A}-\frac{q}{2m}\gamma ^{5}\mathcal A^{0}\right] -im_{r}\beta
,  \label{eq:Weff}
\end{equation}%
with $\hat{\bm\mu }_{\Sigma }=q\bm{\Sigma }/2m$. Equation~(\ref{eq:Weff}) is
the relativistic parent of (\ref{eq:Hcm-intro}): the spatial part of $%
\mathcal A_{\mu }$ couples through the spin operator $\bm\Sigma $, while
the temporal part enters through the odd operator $\gamma ^{5}$. The
nonrelativistic limit completes the comparison with (\ref{eq:Hcm-intro}). Of
the two terms in (\ref{eq:Weff}), the electric contribution is suppressed,
since the matrix element of the odd operator $\gamma ^{5}$ is $O(v/c)$. The
effective block thus reduces to 
\begin{equation}
\mathcal{W}_{\mathrm{eff}}\rightarrow ig\,\hat{\bm\mu }\cdot \left( \mathbf{A%
}_{+}-\mathbf{A}_{-}\right) -im_{r},\qquad \hat{\bm\mu }=\frac{q}{2m}%
\bm{\sigma },  \label{eq:NRlimit}
\end{equation}%
The first term is precisely the lower off-diagonal entry of (\ref%
{eq:Hcm-intro}), whereas the second is the field-independent scalar mixing
already encountered in earlier two-brane realizations. In the regime where the scalar channel is negligible, the familiar Rabi
oscillation driven by $H_{\mathrm{cm}}$ in the phenomenological analyses of
Refs.~\cite{PRD2010,PLB2012,PRD2015,PLB2016,MURMUR} therefore follows
directly. Its explicit form is not needed for the string matching.

\subsection{Gauge covariance}

\label{sec:gaugecovariance}

Under $U(1)_{+}\times U(1)_{-}$ one has $\psi _{\pm }\rightarrow
e^{-iq\alpha _{\pm }}\psi _{\pm }$ and $A_{\mu }^{\pm }\rightarrow A_{\mu
}^{\pm }+\partial _{\mu }\alpha _{\pm }$. Mixed bilinears are not invariant:
they acquire the phase $e^{-iq\theta }$ with $\theta =\alpha _{+}-\alpha _{-}
$, while $\mathcal A_{\mu }\rightarrow \mathcal A_{\mu }+\partial _{\mu
}\theta $. Both sides of (\ref{eq:axialid}) then generate the same extra
term $-iq\,\partial _{\mu }\theta \,J_{5}^{\mu }$, so that the mixed-current
identity is gauge covariant: both sides acquire the same bilocal gauge
phase. An off-diagonal coefficient multiplying the mixed bilinear should
therefore be understood as the gauge-fixed representation of the discrete
link connection in the two-sheeted description \cite{PRD2024,EPJC2025},
rather than as an invariant coefficient under two independent local $U(1)$
transformations. In the string construction below this link structure is
made explicit by the D7 Wilson transporter, together with the
non-perturbative factor compensating the D3 trace charge
(see Sec.~\ref{sec:chargedmodes}).
Thus both descriptions enforce the same relative-gauge requirement, although
its microscopic realization is different.

\subsection{Status of the scalar mixing}

The scalar and pseudoscalar entries in Eq.~(\ref{eq:Dop})
describe distinct transition channels. The spin-dependent
relative-potential structure associated with the pseudoscalar
entry is the target of the present matching. Identifying
this structure does not require the scalar contribution
to be negligible. The relative importance of the two channels
is a separate question involving both the mixing coefficients
and the states participating in the transition.

Previous phenomenological studies have considered a regime in which the magnetic contribution dominates the scalar mixing \cite{PLB2012,PRD2015,PLB2016,MURMUR,IJMPA2019,EPJC2012}. Within the Pauli description used in those studies, the astrophysical estimates of the relative vector potential motivate a large hierarchy between the two coupling strengths. For an elementary charged fermion, the corresponding coefficient ratio is
\begin{equation}
\varrho \equiv
\frac{m_r}{g\,q|\bm{\mathcal A}|/(2m)}
=\frac{m_r}{g}\frac{2m}{q|\bm{\mathcal A}|}.
\label{eq:rho}
\end{equation}
The ratio \(m_r/g\) is model dependent in general. For the single chiral mass entry generated by the minimal E3 sector assumed below, however, its scalar and pseudoscalar decomposition fixes \(m_r=-g\), giving
\begin{equation}
|\varrho|=\frac{2m}{|q|\,|\bm{\mathcal A}|}.
\label{eq:rhosingleE}
\end{equation}

The use of this ratio to infer magnetic dominance requires the nonrelativistic approximation to remain valid for the states being compared. A large vector potential alone does not invalidate that approximation: the relevant momenta are the kinetic momenta \(\bm\pi_\pm=\mathbf p-q\mathbf A_\pm\). Nevertheless, in homogeneous spatial backgrounds with constant mixing, the two plane-wave states coupled by the mixing have the same canonical momentum. Hence \(\bm\pi_+-\bm\pi_-=-q\bm{\mathcal A}\) and \(|q|\,|\bm{\mathcal A}|\leq|\bm\pi_+|+|\bm\pi_-|\). If both states are nonrelativistic, the right-hand side is much smaller than \(2m\). In this case, the condition \(|q|\,|\bm{\mathcal A}|/(2m)\gg1\) cannot be reached within the assumed approximation. The coefficient ratio therefore does not by itself establish a controlled magnetic-dominated regime for the elementary charged fermions considered here.

This restriction should be distinguished from the effective neutron phenomenology of Refs.~\cite{EPJC2012,IJMPA2019}. A neutron has a non-zero magnetic moment but no minimal electric charge, so the charged-particle momentum argument above does not apply directly to its centre-of-mass motion. The derivation and validity of its composite transition operator involve additional assumptions, which are not addressed by the present elementary-fermion matching. For a specified physical system, the relative transition strengths must be assessed using the appropriate states and matrix elements.

The scalar companion is therefore retained in this work and its presence does not remove the pseudoscalar operator or its relative-gauge structure. The following sections determine how these structures arise from the assumed E3 sector and D7 gauge transport and the magnetic dominance is not required for that operator matching.

\section{Minimal D3--D7--D3 geometry and signed holomorphic masses}

\label{sec:string}

To connect the local string mechanism to the two-sheeted description, 
we seek to reproduce the target amplitude in Eq.~(\ref{eq:spinoramp}) for an 
elementary charged fermion. For this purpose, we consider a symmetric D3--D7--D3 
configuration in which the two D3--D7 matter sectors have equal physical masses 
and identical D7 charge, while their opposite transverse separations give rise 
to a relative sign between their holomorphic masses. Since perturbative off-diagonal 
bilinears are absent, we investigate instead the chiral cross-mass that 
an E3 instanton would generate under the charged-spectrum assumptions 
specified below, together with the associated D7 gauge transport.
Realizing this mechanism requires more than the smooth brane configuration 
alone, which does not supply the assumed charged spectrum 
(see Sec.~\ref{sec:chargedmodes}). A compatible local projection must therefore 
be specified, followed by an embedding into a globally consistent compactification. These 
are distinct model-building steps along the roadmap outlined in Sec.~\ref{sec:roadmap}. Throughout 
the present analysis, the fermions are treated as elementary charged fields, 
with no constituent mass, hadronic matrix element, or virtual direct D3--D3 mediator introduced.

\paragraph*{Notation for non-string readers.}
The objects $D3_\pm$ and $D7$ are space-filling D-branes, whereas an E3 is a 
Euclidean D3-brane wrapping an internal four-cycle and therefore acts as an 
instanton in the four-dimensional effective theory. The traditional 
shorthand ``37'' (``73'') denotes an oriented open string with one endpoint 
on a D3 and the other on the D7. The light modes of such strings are ordinary 
four-dimensional fields. Except where the conventional sector name is useful 
in an appendix, we write D3--D7 explicitly in the main text. We denote the 
two $\mathcal{N}=1$ chiral superfields in this D3--D7 hypermultiplet by $Q_s$ 
and $\widetilde Q_s$ (where $s=\pm$ is the brane label). By contrast, the 
symbols $\lambda_\pm,\eta,\widetilde\eta$ introduced below are fermionic E3 
zero modes: they are Grassmann collective coordinates of the instanton, not 
additional propagating four-dimensional particles. Chan--Paton labels are 
simply the endpoint/gauge labels carried by an open string. Requiring those 
labels to close around a world-sheet boundary is the microscopic version of the 
corresponding gauge-charge selection rule. Detailed Chan--Paton matrix notation 
is therefore kept in Appendix~\ref{app:CFT}. In the main text we specify 
open-string orientations directly by their endpoints.

\subsection{Complex coordinates and brane positions}

\begin{figure*}[t]
  \centering
  \includegraphics[width=0.90\textwidth]{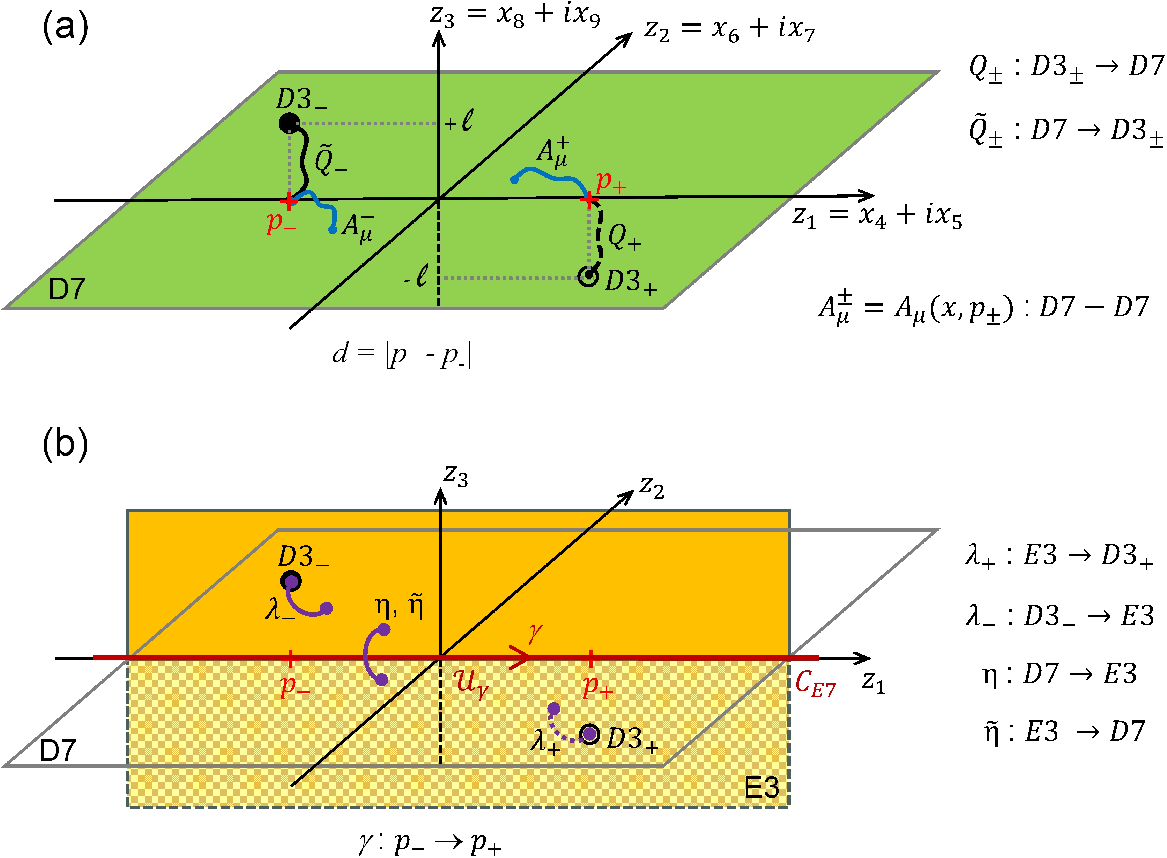} \vspace*{8mm}
	\caption{Local D3$_+$--D7--D3$_-$/E3 geometry under study. (a) D3$_\pm$ sit at
$z_3=\mp\ell$ and project to the D7 points $p_\pm$, separated by $d$ along
$z_1$. $A_\mu^\pm\equiv A_\mu(x,p_\pm)$ are values of the same D7
connection. (b) The E3 wraps $z_2=0$ and intersects the D7 along $C_{E7}$,
which contains $p_\pm$. The charged zero modes are shown with the
orientations required in Eq.~(\ref{eq:minzeromodes}). For D3-branes at
smooth points the E3--D3 modes are vector-like, and this chirality requires
a projection (see Sec.~\ref{sec:chargedmodes}).
$\gamma:p_-\to p_+\subset C_{E7}$ is the reference path for the D7
Wilson transporter.}
  \label{fig:localgeometry}
\end{figure*}

We work locally in Type IIB string theory on $\mathcal{M}^{1,3}\times X_6$,
where $X_6$ denotes the internal six-dimensional space, and choose complex coordinates 
\begin{equation}
z_1=x^4+i x^5,\qquad z_2=x^6+i x^7,\qquad z_3=x^8+i x^9.  \label{eq:zcoords}
\end{equation}
The D7 occupies the local holomorphic divisor
\begin{equation}
D7:\quad z_3=0,  \label{eq:D7divisor}
\end{equation}
so it fills $\mathcal{M}^{1,3}$ and the $z_1,z_2$ directions. Take $d>0$ and 
$\ell>0$ and place the two parallel D3s at 
\begin{align}
D3_+:&\quad (z_1,z_2,z_3)=\left(+\frac d2,0,-\ell\right),  \notag \\
D3_-:&\quad (z_1,z_2,z_3)=\left(-\frac d2,0,+\ell\right).
\label{eq:branePositions}
\end{align}
Their projections onto the D7 are therefore the two points 
\begin{equation}
p_+:z_1=+\frac d2,\qquad p_-:z_1=-\frac d2,  \label{eq:pmp}
\end{equation}
with $z_2=z_3=0$. The corresponding D3--D7 geometry, including the two
endpoint loci $p_\pm$ and the equal DD stretches $\ell$, is shown in Fig.~%
\ref{fig:localgeometry}(a).

The brane array is 
\begin{equation}
\begin{array}{c|cccccccccc}
& 0 & 1 & 2 & 3 & 4 & 5 & 6 & 7 & 8 & 9 \\ \hline
D3_\pm & \times & \times & \times & \times & - & - & - & - & - & - \\ 
D7 & \times & \times & \times & \times & \times & \times & \times & \times
& - & -%
\end{array}%
.  \label{eq:D3D7array}
\end{equation}
For each $s=\pm$, an open string stretched between $D3_s$ and the D7 has four 
ND directions, meaning directions tangent to one brane and transverse to the 
other. The two branes are also separated by $|z_{3,s}|=\ell$ in the common DD 
plane $z_3$ \cite{Polchinski:1998rq}. The physical ground-state mass is therefore
\begin{equation}
m\equiv\frac{\ell}{2\pi\alpha^{\prime}}.  \label{eq:D3D7mass}
\end{equation}
Direct strings stretched between the two D3 branes are spectators in the construction below. No
expectation value or virtual exchange of a D3$_+$--D3$_-$ field is used.

\subsection{The D3--D7 hypermultiplet and the sign of its holomorphic mass}

\label{sec:holmass}

For each D3 brane the two oppositely oriented D3--D7 open strings form an $\mathcal{N}=2$ 
hypermultiplet in the four-dimensional low-energy theory. Written in $\mathcal{N}=1$ language, 
we denote the two chiral superfields by \footnote{In chiral superspace coordinates
$y^\mu=x^\mu+i\theta\sigma^\mu\bar\theta$, a chiral superfield has the
component expansion
$Q_s(y,\theta)=\phi_s(y)+\sqrt{2}\,\theta\chi_s(y)+\theta^2F_s(y)$,
and similarly for $\widetilde Q_s$. Here $\phi_s$ is the complex scalar
component, $\chi_s$ the left-handed Weyl fermion, $F_s$ a non-propagating
auxiliary complex scalar, and $\theta$ the anticommuting Grassmann
coordinate of $\mathcal{N}=1$ superspace. The latter should not be confused
with the Grassmann collective coordinates of the E3 instanton introduced
below.}
\begin{equation}
Q_s:\ D3_s\to D7,\qquad \widetilde Q_s:\ D7\to D3_s,\qquad s=\pm.  \label{eq:QCP}
\end{equation}
Thus $Q_s$ and $\widetilde Q_s$ are four-dimensional matter superfields arising 
from open-string ground states. They are not branes or internal coordinates. If $\chi_s$ 
and $\widetilde\chi_s$ denote their Weyl-fermion components, the corresponding four-component Dirac fermion is
\begin{equation}
\Psi_s\equiv 
\begin{pmatrix}
\chi_s \\ 
\widetilde\chi_s^\dagger%
\end{pmatrix}%
.  \label{eq:D3D7Dirac}
\end{equation}
Throughout, a dagger on a two-component Weyl field denotes Hermitian
conjugation, whereas an overbar on a four-component spinor denotes the
Dirac adjoint, $\bar\Psi\equiv\Psi^\dagger\gamma^0$.
At energies below the string scale this open-string sector is described by a 
four-dimensional supersymmetric effective field theory. It is therefore natural 
to write its interactions in terms of an $\mathcal{N}=1$ superpotential. Then, the relevant D3--D7 coupling is
\begin{equation}
W_{D3D7}=\widetilde Q_s\left(\Phi_7^3-\Phi_s^3\right)Q_s,
\label{eq:D3D7superpotential}
\end{equation}
where $\Phi_s^3$ and $\Phi_7^3$ are adjoint chiral superfields whose scalar 
expectation values encode, respectively, the position of $D3_s$ and the transverse 
position of the D7 in the complex $z_3$ plane \cite{CamaraIbanezUranga}. We use the conventional normalization
\begin{equation}
\left\langle\Phi_7^3-\Phi_s^3\right\rangle
=\frac{z_{3,7}-z_{3,s}}{2\pi\alpha^{\prime}},
\label{eq:PhiPosition}
\end{equation}
so Eq.~(\ref{eq:D3D7superpotential}) simply says that the relative brane 
separation becomes the holomorphic mass of the D3--D7 matter multiplet. No independent 
dynamical role for the fluctuations of $\Phi_s^3$ or $\Phi_7^3$ is used below.

We choose the D7 position as the origin and the real $z_3$ axis as in Eq.~(\ref%
{eq:branePositions}). Then the two holomorphic mass parameters are 
\begin{equation}
m_+=+m,\qquad m_-=-m,  \label{eq:signedmasses}
\end{equation}
although both physical spectra have mass $|m_s|=m$. The opposite signs
reflect the opposite orientations of the two $z_3$ separations displayed in
Fig.~\ref{fig:localgeometry}(a), while their absolute lengths are equal.
Hence the diagonal superpotential is 
\begin{equation}
W_{\mathrm{diag}}=m Q_+\widetilde Q_+-m Q_-\widetilde Q_-.  \label{eq:Wdiag}
\end{equation}
The sign in Eq.~(\ref{eq:Wdiag}) is not a sign of a negative physical
energy. It is the phase of a holomorphic Dirac mass in the canonical
geometric $\mathcal{N}=1$ basis fixed by Eq.~(\ref{eq:D3D7superpotential}) and the
chosen $z_3$ orientation. By itself this phase is redefinition dependent.
In particular, with the single instanton-generated off-diagonal entry
considered below, it can be reversed by the rephasing $Q_-\to-Q_-$, which
leaves $Q_+\widetilde Q_-$ untouched (see Sec.~\ref{sec:gamma5}). We nevertheless
keep the natural geometric basis, in which the diagonal masses and the
off-diagonal coefficient are written in the same string conventions. This
makes the basis conversion of Sec.~\ref{sec:gamma5} explicit.

\subsection{Perturbative Chan--Paton obstruction}

\label{sec:CPobstruction}

An off-diagonal mass term --- responsible for visible--hidden sector mixing --- 
would contain either $Q_+\widetilde Q_-$ or $Q_-\widetilde Q_+$. Let us consider the first 
bilinear. The field $Q_+$ comes from a string $D3_+\to D7$, whereas $\widetilde Q_-$ comes 
from $D7\to D3_-$. Their two endpoints therefore do not form a closed boundary 
sequence: after passing through the D7, the open-string boundary starts on $D3_+$ 
and ends on the distinct brane $D3_-$. Equivalently, $Q_+\widetilde Q_-$ has 
charge $(+1,-1)$ under $U(1)_{3+}\times U(1)_{3-}$ and is neutral 
under $U(1)_7$. Hence a perturbative disk with only these two matter 
insertions cannot generate the desired off-diagonal mass. The identical 
argument applies to $Q_-\widetilde Q_+$. This selection rule can equivalently 
be expressed in terms of Chan--Paton matrices, as shown in Appendix~\ref{app:CFT}. This 
statement does not exclude higher-point disks with additional insertions in an
enlarged model: However, such a route is outside the minimal construction considered here.

By contrast, a non-perturbative E3-instanton vertex can evade this
obstruction, provided that its full contribution remains gauge invariant.
This requires the instanton weight multiplying $Q_+\widetilde Q_-$
to carry charge $(-1,+1)$ under $U(1)_{3+}\times U(1)_{3-}$.
For this compensation to occur, the gauged RR axions must enter the
E3 action with coefficients that produce the required transformation
of the instanton weight. The presence of a
Green--Schwarz/St\"uckelberg coupling alone does not guarantee
this condition. In a consistent background, these coefficients agree with the net
charged zero-mode indices. This agreement fixes the net charge
required by gauge invariance, but does not determine the multiplicities
of additional vector-like zero-mode pairs. Such pairs must therefore
be accounted for separately when specifying the charged spectrum
(see Sec.~\ref{sec:chargedmodes}).

To illustrate how the required charge compensation can arise,
Appendix~\ref{app:chirality} contrasts two settings.
We first consider the smooth brane array in Appendix~\ref{app:smooth} and explain why it does not supply the
required instanton charge. We then turn, in Appendix~\ref{app:orbifold}, to an illustrative example involving
fractional D3 sectors at a common orbifold singularity, where this
charge compensation can be realized
\cite{IbanezRabadanUrangaU1,IbanezUrangaInstanton}.
This example thus illustrates how the charged ingredients of the
mechanism can arise, while the construction of the full configuration
studied here remains a separate model-building task.

To interpret this example correctly, it is also important to
distinguish the D3 trace $U(1)$ factors from electromagnetism.
Indeed, the instanton is neutral under the diagonal D7 $U(1)$.
Nevertheless, its transformation under the individual D7-node
factors must also be included before any recombination.

\section{Conditional E3-instanton mechanism for the chiral off-diagonal mass terms}

\label{sec:instanton}

The perturbative Chan--Paton obstruction of Sec.~\ref{sec:CPobstruction}
motivates a non-perturbative source for the off-diagonal mass term. We now
ask whether an E3 sector can supply the charged zero modes needed to close
the corresponding Chan--Paton chains and generate the forbidden bilinear.
Two conditions will be kept explicit throughout: the required fermionic
zero modes must survive all compactification/string projections, and the
full world-sheet disk amplitudes that couple them to the D3--D7 matter
fields must be non-zero.

\subsection{Local E3-instanton candidate}

\label{sec:E3candidate}

Let us now introduce the non-perturbative E3-instanton used in the construction. An E3 is a 
Euclidean D3-brane: unlike the space-filling $D3_\pm$, it is pointlike in four-dimensional 
spacetime and wraps four internal real directions, so a finite-action E3 configuration 
contributes as an instanton to the four-dimensional effective action. We consider an E3 
whose local holomorphic divisor (a complex-codimension-one submanifold of the internal space) is
\begin{equation}
E3:\qquad z_2=0,  \label{eq:E3divisor}
\end{equation}
so that it wraps the $z_1,z_3$ directions and is pointlike in $\mathcal{M}%
^{1,3}$. Locally, 
\begin{equation}
\begin{array}{c|cccccccccc}
& 0 & 1 & 2 & 3 & 4 & 5 & 6 & 7 & 8 & 9 \\ \hline
D3_\pm & \times & \times & \times & \times & - & - & - & - & - & - \\ 
D7 & \times & \times & \times & \times & \times & \times & \times & \times
& - & - \\ 
E3 & - & - & - & - & \times & \times & - & - & \times & \times%
\end{array}%
.  \label{eq:E3array}
\end{equation}
The E3 passes through both D3 positions because both have $z_2=0$. Its
intersection with the D7 is the complex curve 
\begin{equation}
C_{E7}=E3\cap D7:\qquad z_2=z_3=0,  \label{eq:E7curve}
\end{equation}
parameterized locally by $z_1$ and therefore containing the projected points 
$p_\pm$. The E3 support and its intersection curve $C_{E7}$, containing the
two projected loci $p_\pm$, are shown in Fig.~\ref{fig:localgeometry}(b).
The highlighted segment will later serve as the reference path $\gamma$ for
the D7 gauge transporter.

The D7 and E3 wrap holomorphic divisors and the D3s are points of the
internal complex manifold, so the local array admits a supersymmetric
Bogomol'nyi--Prasad--Sommerfield (BPS) alignment of the branes of the type
used for E3 superpotential effects
\cite{BlumenhagenCveticWeigand2007,IbanezUrangaMajorana,FloreaKachruMcGreevySaulina,IbanezUrangaInstanton,BCKW,BianchiCollinucciMartucci}.
Here ``BPS'' means only that the local brane orientations can preserve
a common supersymmetry. Extending this local alignment to a compact
model requires satisfying the global supersymmetry
($\kappa$-symmetry), orientifold-projection, and world-volume flux
conditions. The orientifold projection relates branes to their image
branes and truncates the open-string spectrum.

Turning to the instanton zero modes, open strings joining the E3 to either a D3 or the D7 have eight mixed
ND directions. In the standard open-string spectrum this removes massless
bosonic ground states in the Neveu--Schwarz (NS) sector but allows fermionic
states in the Ramond (R) sector. Such states can become fermionic collective coordinates of the
instanton, as detailed by the explicit world-sheet counting
presented in Appendix~\ref{app:CFT}.
For a D3-brane at a smooth point of the instanton divisor, nothing
distinguishes the two orientations of the E3--D3 string: each provides one
fermionic zero mode, so that the E3--D3$_s$ sector contains a vector-like
pair, coupled through the transverse separation of D3$_s$ from the E3
\cite{Ganor}. The chiral content required below therefore does not follow
from the local array of Fig.~\ref{fig:localgeometry} alone: it requires a
projection, discussed in Sec.~\ref{sec:chargedmodes}.
This local counting determines only the possible
type of zero mode. The actual number of independent zero modes in four
dimensions is fixed only after the compactification and its projections are
specified.

A finite instanton action also requires the local divisor $z_2=0$ in
Eq.~(\ref{eq:E3divisor}) to arise as a local patch of a compact
holomorphic divisor $\Sigma_E\subset X_6$, which is the global internal
four-cycle wrapped by the E3. We denote this instanton configuration by
$E$. We assume an orientifold-invariant $O(1)$ E3 instanton. The notation $O(1)$ means that the
orientifold leaves only the discrete Chan--Paton group $O(1)$ on the
instanton, rather than a continuous instanton $U(1)$. It does not introduce a
new four-dimensional gauge interaction. The $O(1)$ projection retains the two universal fermionic modes associated
with $d^2\theta$. We additionally assume that all other neutral fermionic
zero modes are absent or lifted, so that the neutral measure is
$d^4x\,d^2\theta$. This includes modes associated with deformations of the
wrapped cycle, internal Wilson lines, or instanton flux. These are conditions
on the divisor, projection and compactification data \cite{BCKW,BianchiCollinucciMartucci,GrimmFluxedInstantons}. Instanton contributions to multi-fermion or higher-derivative F-terms
\cite{BeasleyWittenHigherF} are outside the scope of the present analysis.

\subsection{Minimal charged zero modes and the two disks}

\label{sec:chargedmodes}

For the two-disk mechanism considered here, we assume exactly four
charged fermionic integration variables,
\begin{align}
\lambda_+:&\ E3\to D3_+, & \eta:&\ D7\to E3, \notag \\
\widetilde\eta:&\ E3\to D7, & \lambda_-:&\ D3_-\to E3.
\label{eq:minzeromodes}
\end{align}
The arrows indicate which branes each open string connects and
specify its orientation. The associated symbols
$\lambda_+,\eta,\widetilde\eta,\lambda_-$ denote Grassmann integration
variables in the instanton measure, rather than ordinary
four-dimensional propagating fields. These variables are referred
to as charged zero modes because they carry charges under the
gauge groups living on the space-filling D3 and D7 branes,
as displayed in Table~\ref{tab:zeromodecharges}.
The designation ``charged'' thus refers to these D3 and D7 gauge
groups, while the E3 itself does not provide an additional
four-dimensional gauge field. 

The assumed charged spectrum must contain no additional unsaturated
modes, since integrating over any such mode would make the instanton
contribution vanish unless further couplings or insertions absorb it.
Appendix~\ref{app:orbifold} provides an illustrative example in a
separate orbifold setting, where the E3--D3 modes and E3--D7
representations can be derived explicitly.
This example does not establish the complete minimal spectrum
assumed here or provide a full realization of the proposed
configuration. In particular, the E3--D7 representations do not fix the zero-mode
multiplicities along the intersection curve, which also depend
on boundary conditions or compact bundle data. Accordingly, Appendix~\ref{app:orientifoldscope} specifies the scope
of this illustration and the orientifold conditions relevant
to the assumed spectrum.

\begin{table}[tbp]
\caption{Open-string origin and gauge charges of the fields and 
fermionic zero modes entering the instanton contribution to $Q_+\widetilde Q_-$. The E3 itself is not a four-dimensional gauge sector.}
\label{tab:zeromodecharges}\centering
\begin{tabular}{ccccc}
\toprule State & Open-string endpoints & $U(1)_{3+}$ & $U(1)_{3-}$ & $U(1)_7$ \\ 
\midrule $\lambda_+$ & $E3\to D3_+$ & $-1$ & $0$ & $0$ \\ 
$Q_+$ & $D3_+\to D7$ & $+1$ & $0$ & $-1$ \\ 
$\eta$ & $D7\to E3$ & $0$ & $0$ & $+1$ \\ 
$\widetilde\eta$ & $E3\to D7$ & $0$ & $0$ & $-1$ \\ 
$\widetilde Q_-$ & $D7\to D3_-$ & $0$ & $-1$ & $+1$ \\ 
$\lambda_-$ & $D3_-\to E3$ & $0$ & $+1$ & $0$ \\ 
\bottomrule &  &  &  & 
\end{tabular}%
\end{table}

The required interaction can be generated by two tree-level open-string
world-sheet disk amplitudes. A disk is the leading open-string world-sheet
with a boundary. Here that boundary is divided into segments lying on the E3,
a D3, and the D7, and the three insertions occur where the boundary condition
changes from one brane to the next. In the standard D-brane instanton
calculus, charged fermionic zero modes couple to open-string matter through
disk amplitudes containing two charged zero-mode insertions and one
matter-field insertion
\cite{CveticRichterWeigand,IbanezUrangaInstanton}.
We denote the two complete disk coefficients by $c_+$ and $c_-$:
\begin{align}
\left\langle V_{\lambda_+}V_{Q_+}V_{\eta}\right\rangle_D
&\equiv c_+\,\lambda_+Q_+\eta,
\notag \\
\left\langle
V_{\widetilde\eta}V_{\widetilde Q_-}V_{\lambda_-}
\right\rangle_D
&\equiv
c_-\,\widetilde\eta\widetilde Q_-\lambda_- .
\label{eq:E3disks}
\end{align}
Here $V_X$ is the world-sheet vertex operator representing the state $X$, 
whereas $c_\pm$ are ordinary complex coefficients. By definition, $c_\pm$ include 
all physical-state and compactification projections, Chan--Paton factors and compact 
wavefunction overlaps. The technical world-sheet superghost (``picture'') assignment 
and the Chan--Paton endpoint checks are given in Appendix~\ref{app:CFT}.

\begin{figure}[t]
  \centering
  \includegraphics[width=0.90\columnwidth]{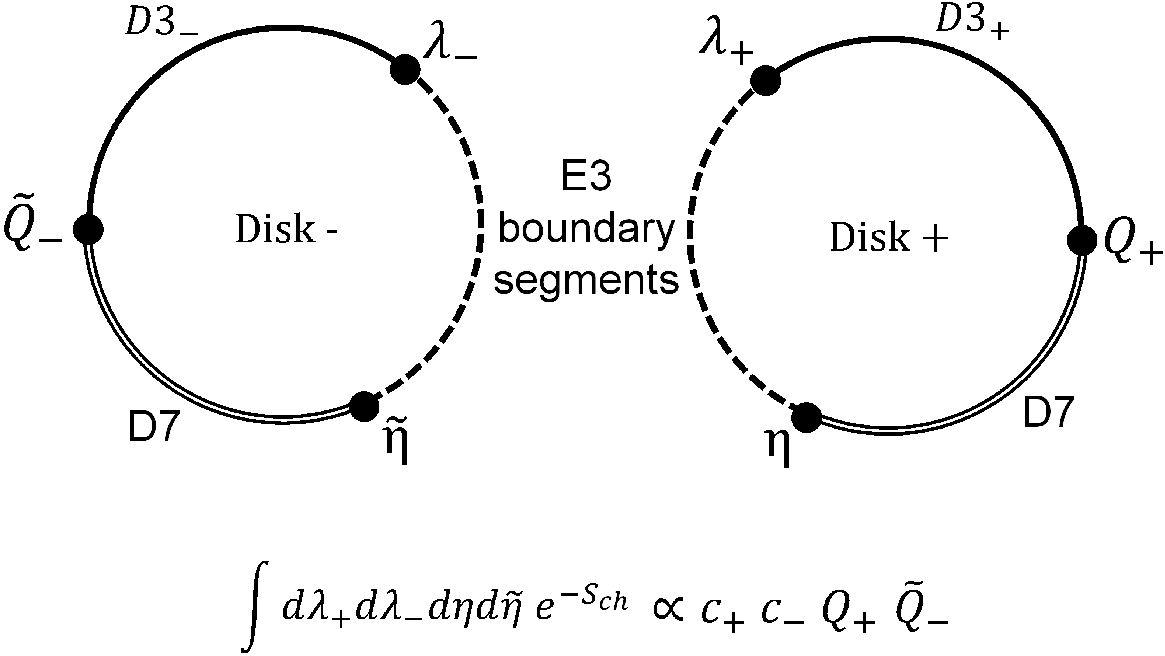} \vspace*{3mm}
  \caption{Charged disks of the E3 instanton configuration $E$ generating
$Q_+\widetilde Q_-$. The four E3--D3 and E3--D7 zero modes are independent
Grassmann variables. Simultaneous saturation of the two disks yields
$c_+c_-Q_+\widetilde Q_-$.}
  \label{fig:process}
\end{figure}

The charged-zero-mode structure of these two disks is summarized schematically in 
Fig.~\ref{fig:process}. Both disk couplings belong to the same E3 instanton configuration $E$.

The matter vertex on each charged disk corresponds to the scalar
component of the relevant D3--D7 chiral multiplet.
These disk couplings determine the induced superpotential,
whose second derivatives yield the mass terms for the
D3--D7 fermions. At finite D3--D7 separation the required realization condition is simply
\begin{equation}
c_+\neq0,\qquad c_-\neq0.
\label{eq:fullDiskCondition}
\end{equation}
This condition includes the Gliozzi--Scherk--Olive (GSO) physical-state projection, 
the remaining world-sheet spin/correlation factors, Chan--Paton endpoint factors, 
and compact internal wavefunction overlaps. Equation~(\ref{eq:E3disks}) is therefore 
schematic notation for the corresponding Wilsonian holomorphic operator, not a claim 
that the complete on-shell string correlator has been evaluated.

\paragraph*{Finite-separation boundary conformal field theory (BCFT) factor.}

Since the D3--D7 matter states are stretched by a finite distance
$\ell$ in the $x^8=\operatorname{Re}z_3$ direction, one must check that
this separation does not by itself force the charged disk amplitudes
to vanish. The $x^8$ free-boson sector is the part of the world-sheet
BCFT directly sensitive to this separation. Appendix~\ref{app:finiteBCFT}
isolates this factor and shows that it is non-zero for every finite
$\ell$ in a fixed T-dual Dirichlet/Wilson-line sector. Denoting this
single factor by $F_s^{(8)}(\ell)$ for $s=\pm$,
\begin{equation}
F_s^{(8)}(\ell)\neq0\qquad (s=\pm,\ |\ell|<\infty).
\label{eq:AfiniteNonzero}
\end{equation}
The remaining factors entering $c_\pm$ are not evaluated here, so the 
full condition remains Eq.~(\ref{eq:fullDiskCondition}).

The endpoint sequence closes on each disk: $E3\to D3_+\to D7\to E3$ for the 
first amplitude and $E3\to D7\to D3_-\to E3$ for the second. This is the 
Chan--Paton selection rule in a notation that makes the geometry explicit. Its 
matrix form is given in Appendix~\ref{app:CFT}. Conditional on (\ref{eq:fullDiskCondition}), the charged-mode action has the form
\begin{equation}
S_{\mathrm{ch}}=c_+\lambda_+Q_+\eta+c_-\widetilde\eta\widetilde Q_-\lambda_-.
\label{eq:Sch}
\end{equation}

Integrating over the charged fermionic zero modes saturates the two
disk couplings in \(S_{\rm ch}\) and gives
\begin{equation}
\int d\lambda_+\,d\lambda_-\,d\eta\,d\widetilde\eta\,
e^{-S_{\rm ch}}
\mkern5mu\propto\mkern5mu
c_+c_-\,Q_+\widetilde Q_- .
\label{eq:Grassmann}
\end{equation}
Thus, Eq.~\eqref{eq:Grassmann} fixes the matter-field structure
generated by the instanton.  The overall field-independent
normalization, including the sign associated with the ordering of the
Grassmann measure, may be absorbed into the instanton prefactor.  This
is the standard saturation of charged instanton zero modes by disk
couplings with matter-field insertions
\cite{BlumenhagenCveticWeigand2007,CveticRichterWeigand,BCKW,IbanezUrangaInstanton}.

Under the assumed \(O(1)\) projection and absence or lifting of all
additional neutral fermionic zero modes, the neutral measure is
\(d^4x\,d^2\theta\), so Eq.~\eqref{eq:Grassmann} can contribute
to the four-dimensional superpotential.  Restoring the classical
instanton weight and the one-loop fluctuation prefactor, the complete
semiclassical contribution is
\begin{equation}
\begin{aligned}
W_E
&=
M_{\mathrm{s}}\,Z_E^{\mathrm{1-loop}}\,e^{-S_E^{\mathrm{cl}}}
\int d\lambda_+\,d\lambda_-\,d\eta\,d\widetilde\eta\,
e^{-S_{\rm ch}}
\\
&=
\mu_{+-}^{\rm hol}\,Q_+\widetilde Q_- ,
\qquad
\mu_{+-}^{\rm hol}
\equiv
M_{\mathrm{s}}\,Z_E^{\mathrm{1-loop}}\,e^{-S_E^{\mathrm{cl}}}\,c_+c_- .
\end{aligned}
\label{eq:muplusminus}
\end{equation}
Here $Z_E^{\mathrm{1-loop}}$ denotes the holomorphic one-loop
fluctuation prefactor with the zero modes removed and $S_E^{\mathrm{cl}}$ is 
the classical Euclidean action of the E3
instanton configuration $E$, wrapping the compact divisor $\Sigma_E$.
In the normalization used here, the factor $M_{\mathrm{s}}\equiv1/\sqrt{\alpha^{\prime}}$ 
supplies the string mass scale required for the coefficient of the bilinear operator.
The decomposition into charged disk amplitudes, the classical
instanton action and the one-loop vacuum amplitude follows the
standard D-brane instanton calculus
\cite{BlumenhagenCveticWeigand2007,BCKW,IbanezUrangaInstanton}. Canonical normalization 
of the D3--D7 fields then converts $\mu_{+-}^{\mathrm{hol}}$ into the coefficient entering
the physical fermion mass matrix. We denote that canonically normalized
coefficient simply by $\mu_{+-}$. At the order retained, we assume no kinetic mixing
between the two D3 sectors. Note that comparable off-diagonal
K\"ahler corrections would require a matrix normalization before extracting
the physical scalar and pseudoscalar entries.

Thus, the microscopic string input controlling the interbrane transition is encoded
in $\mu_{+-}^{\rm hol}=M_{\mathrm{s}}Z_E^{\mathrm{1-loop}}e^{-S_E^{\mathrm{cl}}}c_+c_-$.
As shown in Sec.~\ref{sec:gamma5}, after canonical normalization $\mu_{+-}$
fixes $g=\mu_{+-}/2$ and $m_r=-\mu_{+-}/2$. The gauge-dependent matrix element
derived below should
therefore be understood as the low-energy matching of this string-generated
operator, rather than as an independent determination of its overall strength.

\paragraph*{Instanton charge and chirality of the E3--D3 zero modes.}
Note that the operator $Q_+\widetilde Q_-$ carries charge $(+1,-1,0)$ under
$U(1)_{3+}\times U(1)_{3-}\times U(1)_7$, whereas the zero modes of
Eq.~(\ref{eq:minzeromodes}) carry the total charge
$Q(E)=(-1,+1,0)$ (see Table~\ref{tab:zeromodecharges}). Gauge invariance of
Eq.~(\ref{eq:muplusminus}) therefore requires $e^{-S_E^{\mathrm{cl}}}$ to
transform with the charge $Q(E)$, in accordance with the general
charge-conservation rule of D-brane instanton calculus \cite{BCKW},
\begin{equation}
\sum_{\Phi}q_a(\Phi)=-Q_a(E),\qquad
Q_a(E)=\sum_{\zeta}q_a(\zeta),
\label{eq:chargematching}
\end{equation}
Here $a=3+,3-,7$ labels the three gauge factors, and $q_a(X)$
denotes the charge of $X$ under $U(1)_a$. The first sum runs
over the matter fields in the generated operator, while the
second runs over the charged fermionic zero modes $\zeta$.
The quantity $Q_a(E)$ is the charge with which the instanton
weight $e^{-S_E^{\mathrm{cl}}}$ must transform.
This transformation is carried by the axionic part of
$S_E^{\mathrm{cl}}$, i.e.\ by a four-dimensional shift-symmetric scalar
descending from the Ramond--Ramond (RR) sector of the compactification, not
specifically the QCD axion.

For clarity, let us choose real D3 gauge parameters $\alpha_\pm$ and the charge
convention $\Phi\mapsto e^{i\sum_a q_a\alpha_a}\Phi$. Then
\begin{align}
Q_+\widetilde Q_-&\mapsto
 e^{i(\alpha_+-\alpha_-)}Q_+\widetilde Q_-,\notag\\
\delta S_E^{\mathrm{cl}}&=i(\alpha_+-\alpha_-),\notag\\
e^{-S_E^{\mathrm{cl}}}&\mapsto
 e^{-i(\alpha_+-\alpha_-)}e^{-S_E^{\mathrm{cl}}}.
\label{eq:physicalEshift}
\end{align}
The transformation of the charged zero-mode measure provides
an additional consistency check of gauge invariance. Since a
Grassmann differential transforms inversely to its variable,
\begin{equation}
\begin{aligned}
d\lambda_+\,d\lambda_-\,d\eta\,d\widetilde\eta
\ \mapsto{}& e^{i(\alpha_+-\alpha_-)}\\
&\!\!\times d\lambda_+\,d\lambda_-\,d\eta\,d\widetilde\eta .
\end{aligned}
\label{eq:chargedMeasureShift}
\end{equation}
Thus the weight cancels the measure phase as well as the charge of the
saturated operator. These equations state the transformation required for
the proposed spectrum and its RR-coupling derivation in the orbifold example
is given in Appendix~\ref{app:orbifold}.

The compensating charge of the instanton weight must follow from
the RR couplings of the E3 and the space-filling branes. A
St\"uckelberg mass for the D3 trace vectors does not by itself
determine this charge, because the axions responsible for that
mass must also couple to the E3 with the appropriate coefficients.

To express how the RR axion couplings determine the charge
of the instanton weight, let $a^I$ denote a basis of real RR axions. The axionic part of the E3 action follows from its
Chern--Simons couplings, while the corresponding couplings of
the space-filling branes determine the gauged axion shifts
\cite{BCKW,IbanezUrangaInstanton}. In a fixed normalization,
these two sets of couplings can be written as
\begin{equation}
S_E^{\mathrm{cl}}\supset i\sum_I c_I a^I,\qquad
\delta a^I=\sum_a k^I_a\alpha_a,
\label{eq:axionCouplingData}
\end{equation}
where $c_I$ is the coefficient with which the axion $a^I$
enters the E3 action, and $k^I_a$ is its shift per unit gauge
parameter $\alpha_a$ under $U(1)_a$. These coefficients are
fixed by the brane configuration and its RR couplings.
Substituting the shifts into the action gives
\begin{equation}
\delta S_E^{\mathrm{cl}}
=i\sum_a\left(\sum_I c_I k^I_a\right)\alpha_a.
\label{eq:shiftsaction}
\end{equation}
Since the instanton weight is $e^{-S_E^{\mathrm{cl}}}$,
its charge under $U(1)_a$ is therefore
\begin{equation}
Q_a(E)=-\sum_I c_I k^I_a.
\label{eq:QaE}
\end{equation}
For the proposed operator $Q_+\widetilde Q_-$, gauge invariance
requires $Q_a(E)=(-1,+1,0)$ for $a=(3+,3-,7)$. Equation~(\ref{eq:QaE}) therefore implies
$\sum_I c_I k^I_{3+}=1$,
$\sum_I c_I k^I_{3-}=-1$, and
$\sum_I c_I k^I_7=0$.

The charge obtained from these RR couplings must also agree
with the total charge of the fermionic zero modes in
Eq.~(\ref{eq:chargematching}). This agreement relates the
axionic transformation of the instanton action to the charged
zero-mode spectrum
\cite{BCKW,IbanezUrangaInstanton}.
This matching fixes the net instanton charge, not the complete spectrum:
adding a vector-like pair leaves Eq.~(\ref{eq:chargematching}) unchanged but
can obstruct the minimal saturation. A non-zero charge requires gauging of
an axionic direction coupled to the instanton. The gauge-boson mass matrix also depends on the axion
kinetic terms and on other gauge factors. In the present analysis, we do not infer its eigenvectors from the
charged index alone. If no axion shifts under the relative D3 transformation and no additional charged field
supplies the compensating factor, the isolated bilinear is forbidden.

For a D3-brane at a smooth point of $\Sigma_E$ in the array
considered here, the E3--D3 zero modes form a vector-like pair
(see Sec.~\ref{sec:E3candidate}), and the D3 trace $U(1)$ does not
shift $S_E^{\mathrm{cl}}$. The instanton is therefore neutral
under $U(1)_{3+}\times U(1)_{3-}$ and cannot generate
$Q_+\widetilde Q_-$ (see Appendix~\ref{app:smooth}).

To illustrate how chiral E3--D3 modes and a compensating
instanton charge can arise, Appendix~\ref{app:orbifold}
examines a separate example with fractional D3-branes at
a common orbifold singularity. In this setting, twisted
RR fields couple to the D3 trace $U(1)$s
\cite{IbanezRabadanUrangaU1} and enter the action of an E3
passing through the singularity
\cite{IbanezUrangaInstanton}. The appendix derives the
E3--D3 spectrum and the local E3--D7 representations,
identifies the two allowed Chan--Paton chains, and checks
both local orbifold twisted tadpole cancellation and the
compensating RR charge. These calculations provide an
explicit check of the charged ingredients in that example,
rather than a complete realization of the massive
separated-endpoint configuration. The relation between
the two settings is specified in Appendix~\ref{app:orientifoldscope}.

In the minimal construction considered here, no independent holomorphic source
for $Q_-\widetilde Q_+$ is assumed.\footnote{Such an opposite
holomorphic entry could in principle arise from an additional instanton
sector with the appropriate charged-zero-mode structure. If both chiral
entries are generated with a suitable relative magnitude and phase, their
scalar contributions can cancel while the pseudoscalar $i\gamma^5$
coupling remains. Realizing this possibility requires additional global
compactification and zero-mode conditions and lies beyond the minimal
single-$E$ setup considered here.} The Hermitian conjugate of
Eq.~(\ref{eq:muplusminus}) belongs to the anti-superpotential and supplies the
conjugate interaction required by the four-dimensional Hermitian action, but it
does not generate a second holomorphic mass entry. The physical consequence of
this single chiral entry is derived in the next section.

\section{From the string mass matrix to the TBTS \texorpdfstring{$%
\gamma^5$}{gamma5} entry}

\label{sec:gamma5}

Given the assumed instanton measure and non-vanishing disk coefficients,
the preceding saturation yields a chiral off-diagonal mass coefficient
$\mu_{+-}$. The TBTS description of
Sec.~\ref{sec:fieldtheory} is formulated in terms of two Dirac fermions
with equal positive masses and contains both pseudoscalar and scalar
interbrane entries. The purpose of this section is to connect these two
descriptions. We show that the characteristic $i\gamma^5$ structure need not
be inserted as a special phase of the instanton amplitude: it emerges,
together with its scalar partner, from the chirality of the
instanton-induced holomorphic mass, once the natural string-basis mass matrix
is converted to the equal-positive-mass TBTS Dirac basis.
This is an algebraic matching for the assumed two-fermion effective theory:
It does not derive that massive truncation from the orbifold example.

\subsection{Two-component mass matrix and four-component Dirac form}

After canonical normalization, the diagonal contribution
$W_{\mathrm{diag}}$ (see Eq. \ref{eq:Wdiag}) and the instanton-induced contribution $W_E$ (see Eq. \ref{eq:muplusminus}) combine
into a single mass superpotential. In the natural string basis this can be
written as
\begin{equation}
W_{\mathrm{mass}}^{\mathrm{str}}=
\begin{pmatrix}
Q_+^{\mathrm{str}} & Q_-^{\mathrm{str}}
\end{pmatrix}
\begin{pmatrix}
m & \mu_{+-} \\
0 & -m
\end{pmatrix}
\begin{pmatrix}
\widetilde Q_+^{\mathrm{str}} \\
\widetilde Q_-^{\mathrm{str}}
\end{pmatrix}.
\label{eq:stringmassmatrix}
\end{equation}
The superpotential is the supersymmetric encoding of the mass
interactions: its second derivatives with respect to the chiral
superfields give the corresponding Weyl-fermion mass terms. In particular, 
in this basis the off-diagonal entry corresponds to
$\mu_{+-}Q_+^{\mathrm{str}}\widetilde Q_-^{\mathrm{str}}$.
Combining the Weyl fields into the Dirac spinors defined in
Eq.~(\ref{eq:D3D7Dirac}), using the standard relation between two- and
four-component spinor formalisms \cite{DreinerHaberMartin}, with
$P_{L,R}=(1\mp\gamma^5)/2$, then gives the fermionic mass Lagrangian

\begin{align}
\mathcal{L}_m^{\mathrm{str}}
={}&-m\bar\Psi_+^{\mathrm{str}}\Psi_+^{\mathrm{str}}
+m\bar\Psi_-^{\mathrm{str}}\Psi_-^{\mathrm{str}}
\notag\\
&-\mu_{+-}\bar\Psi_-^{\mathrm{str}}P_L
\Psi_+^{\mathrm{str}}
+\mathrm{h.c.}.
\label{eq:Lstringmass}
\end{align}
The identity $\chi_s\widetilde\chi_t=\bar\Psi_tP_L\Psi_s$ is recorded explicitly
in Appendix~\ref{app:Diracconversion}. Equation~(\ref{eq:Lstringmass}) thus displays explicitly, in the
natural string basis, the relative diagonal-mass sign and the single
chiral off-diagonal mass coefficient before the field redefinition
leading to the positive-mass TBTS basis is performed.

\subsection{Positive-mass TBTS basis}

The TBTS operator of Sec.~\ref{sec:fieldtheory} uses the same positive
Dirac mass $m$ on the two sheets. To convert Eq.~(\ref{eq:Lstringmass})
to this basis, the plus sector is left unchanged, while an axial
redefinition is performed only on the minus-sector Dirac field:
\begin{equation}
\Psi_+^{\mathrm{TBTS}}=\Psi_+^{\mathrm{str}},
\qquad
\Psi_-^{\mathrm{TBTS}}
=i\gamma^5\Psi_-^{\mathrm{str}}.
\label{eq:axialredef}
\end{equation}
At the level of the corresponding $\mathcal{N}=1$ chiral superfields,
this is implemented by
\begin{align}
Q_+^{\mathrm{TBTS}}
&=Q_+^{\mathrm{str}},
&
\widetilde Q_+^{\mathrm{TBTS}}
&=\widetilde Q_+^{\mathrm{str}},
\notag\\
Q_-^{\mathrm{TBTS}}
&=-iQ_-^{\mathrm{str}},
&
\widetilde Q_-^{\mathrm{TBTS}}
&=-i\widetilde Q_-^{\mathrm{str}}.
\label{eq:superfieldrephase}
\end{align}

The ``$+$'' sector is unchanged. In the ``$-$'' sector, the axial
redefinition reverses the sign of the scalar bilinear,
\begin{equation}
\bar\Psi_-^{\mathrm{TBTS}}\Psi_-^{\mathrm{TBTS}}
=
-\bar\Psi_-^{\mathrm{str}}\Psi_-^{\mathrm{str}},
\label{eq:minusmassflip}
\end{equation}
so that the two diagonal mass terms in the TBTS basis both take the
standard positive-mass form
$-m\bar\Psi_s^{\mathrm{TBTS}}\Psi_s^{\mathrm{TBTS}}$.
The vector current, and hence the D7 electric charge, is unchanged by
this redefinition.

From this point onward, $\Psi_\pm$ without a basis superscript denote
the positive-mass TBTS fields. The superscript ``str'' is retained only
when the pre-redefinition string basis must be displayed explicitly.

Using Eq.~(\ref{eq:axialredef}), the off-diagonal term from Eq.~(\ref{eq:Lstringmass}) becomes
\begin{equation}
\mathcal{L}_{\mathrm{off}}^{\mathrm{TBTS}}=
\bar\Psi_-\left(i g\gamma^5+i m_r\right)\Psi_+ +\mathrm{h.c.},
\label{eq:LTBTSmix}
\end{equation}
where a residual relative vector rephasing may be used to make
$\mu_{+-}$ real and positive, and
\begin{equation}
g=\frac{\mu_{+-}}{2},\qquad
m_r=-\frac{\mu_{+-}}{2}=-g.
\label{eq:gmfrommu}
\end{equation}
These are the same $g$ and $m_r$ introduced in
Sec.~\ref{sec:fieldtheory}. Equation~(\ref{eq:gmfrommu}) is the central
algebraic result of the single-instanton construction: the pseudoscalar
entry is not inserted into the E3 amplitude by hand. It appears because the
instanton-induced off-diagonal mass is holomorphic, and hence chiral (it
contains only $P_L$, together with its Hermitian conjugate), whereas the TBTS
basis assigns both Dirac states the same positive, non-chiral mass. A chiral
entry decomposes into scalar and pseudoscalar parts of equal magnitude, which
is the origin of $m_r=-g$. The relative sign of the diagonal holomorphic
masses in the geometric basis plays no physical role in this result. With a
single off-diagonal entry, it can be reversed by the rephasing
$Q_-\to-Q_-$, which leaves $Q_+\widetilde Q_-$ untouched. Starting from equal
positive masses and a real positive chiral coefficient, defining the new
minus-sector field by $\Psi_-^{\mathrm{new}}=i\Psi_-^{\mathrm{old}}$ gives
$-i\mu_{+-}\bar\Psi_-^{\mathrm{new}}P_L\Psi_+$ and hence again
Eqs.~(\ref{eq:LTBTSmix})--(\ref{eq:gmfrommu}).

The relation $|m_r|=|g|$ is not peculiar to the present construction. In the
domain-wall reduction of Ref.~\cite{PRD2010}, $m_r$ and $g$ arise from
closely related overlap integrals and explicit localized modes include the
case $m_r=-g$, whereas the string-inspired phenomenological model of
Ref.~\cite{IJMPA2019} gives the corresponding equality in magnitude. The
scalar term is physically distinct from the spin-dependent swapping term
and cannot in general be discarded. The assumed minimal E3 sector therefore
needs no second holomorphic source to reproduce the pseudoscalar operator
structure, but supplies its scalar companion with equal coefficient magnitude.
Their relative transition strengths depend on the states, as explained in
Sec.~\ref{sec:fieldtheory}.
A final quantum caveat concerns the axial field redefinition in
Eq.~(\ref{eq:axialredef}). It is a change of variables, not an exact axial
symmetry. Its anomalous Jacobian shifts the corresponding CP-odd gauge theta
angles \cite{Fujikawa1979,Fujikawa1980}, which must be tracked in a compact
orientifold. This does not alter the vector coupling and therefore does not affect the
tree-level soft D7 gauge matching developed in the next section.

\section{D7 gauge transport and leading low-energy matching}

\label{sec:relative}

The microscopic strength of the interbrane transition is already encoded in
the canonically normalized E3-induced coefficient $\mu_{+-}$, whose holomorphic
source is given in Eq.~(\ref{eq:muplusminus}), and hence in $g$ and $m_r$. The
purpose of this section is different: to determine the
gauge-covariant D7 realization of this string-generated operator and its leading
soft four-dimensional matrix element. The analysis below is therefore a
low-energy matching calculation, rather than an independent world-sheet
determination of the transition coefficient.
It assumes distinct endpoint loci and a non-zero chiral coefficient in the
same configuration. Their simultaneous microscopic realization is not
supplied by the coincident-node example of Appendix~\ref{app:orbifold}.

\subsection{Wilson completion and the relative D7 connection}

\label{sec:WilsonCompletion}

The two matter insertions entering the instanton-generated bilinear
$Q_+\widetilde Q_-$ attach to the D7 at the distinct internal points $p_+$
and $p_-$. Their endpoint phases therefore do not cancel under an
internal-position-dependent D7 gauge transformation. Gauge covariance
requires a Wilson transporter between the two points
\cite{HillWilsonLine}. Let us choose a reference path
$\gamma$ on the E3--D7 intersection curve from $p_-$ to $p_+$ (see Fig.~%
\ref{fig:localgeometry}(b)).
Let $y^a$ denote internal coordinates on the D7, with $a$ an internal D7 index, and define
\begin{equation}
U_\gamma(x)\equiv \exp\!\left[iq\int_\gamma A_a(x,y)\,dy^a\right].
\label{eq:WilsonTransporter}
\end{equation}
Here $q$ is the D7 charge of the D3--D7 Dirac fermion. For $A_M\to A_M+\partial_M\alpha$,
\begin{equation}
U_\gamma\to e^{iq\alpha(p_+)}U_\gamma e^{-iq\alpha(p_-)}.
\label{eq:Utransform}
\end{equation}
At the component level, endpoint gauge covariance is implemented by
inserting $U_\gamma$ in the instanton-generated off-diagonal fermion
mass term. Allowing in addition for a possible gauge-invariant dressing
of the local E3 vertex, we write
\begin{equation}
\mathcal{L}_{\mathrm{off}}^{\mathrm{str}}[A]=-
\mu_{+-}\bar\Psi_-^{\mathrm{str}}U_\gamma G_E[A]P_L
\Psi_+^{\mathrm{str}}+\mathrm{h.c.}
\label{eq:Lstringmassfunctional}
\end{equation}
The Hermitian conjugate carries the opposite endpoint orientation
and contains $U_\gamma^\dagger=U_\gamma^{-1}$.
Here $U_\gamma$ is required by endpoint gauge covariance, whereas $G_E[A]$
parametrizes additional model-dependent gauge dependence of the local E3
vertex. We normalize $G_E[A^{(0)}]=1$ in the reference background. Its
soft-field behavior is analyzed in Appendix~\ref{app:softVectorDressing}.

The same physical relative connection introduced in
Sec.~\ref{sec:fieldtheory} admits, in the D7 realization, a gauge-invariant
Wilson-completed representation. We therefore retain the same symbol and define
\begin{align}
\mathcal{A}_\mu(\gamma)&\equiv A_\mu(x,p_+)-A_\mu(x,p_-)
-\frac{1}{iq}U_\gamma^{-1}\partial_\mu U_\gamma  \notag \\
&=-\int_\gamma F_{\mu a}(x,y)\,dy^a.
\label{eq:relativeConnection}
\end{align}
In axial gauge $A_a=0$ along $\gamma$ this reduces to the ordinary endpoint difference,
\begin{equation}
\mathcal{A}_\mu(\gamma)=A_\mu(p_+)-A_\mu(p_-).
\label{eq:axialgaugeA}
\end{equation}
Upon identifying $A_\mu^\pm=A_\mu(x,p_\pm)$, Eq.~(\ref{eq:axialgaugeA}) is
precisely the relative connection $\mathcal A_\mu=A_\mu^+-A_\mu^-$ used in
the two-sheeted description. Thus the Wilson transporter does not introduce a
new physical field. It supplies the gauge-invariant completion of the same
relative degree of freedom in the D7 realization. For a fixed reference path
$\gamma$, we henceforth omit the argument $(\gamma)$ and simply write
$\mathcal A_\mu$.

\subsection{Wilson-completed mixed axial identity at leading soft order}

\label{sec:chiralDressing}

The Wilson completion of the instanton-generated mass term is displayed
explicitly in Eq.~(\ref{eq:Lstringmassfunctional}). The additional factor
$G_E[A]$ is not required by endpoint covariance. It parametrizes only a
possible model-dependent scalar dressing of the chiral mass vertex.
To determine the leading soft-vector contribution of the
Wilson-completed mass operator, only the power counting
of $G_E[A]$ is needed. As shown in Appendix~\ref{app:softVectorDressing}, for a four-dimensional D7 vector
fluctuation $\delta A$ of soft momentum $k$,
\[
\delta G_E^{(1)}=O(k\,\delta A),\qquad
\partial_\mu\delta G_E^{(1)}=O(k^2\delta A).
\]
Hence the leading $O(k^0\delta A)$ contribution of the
Wilson-completed mass operator is unchanged by the scalar
dressing $G_E[A]$. The scope of this soft-momentum expansion
is discussed in Appendix~\ref{app:softVectorDressing}.

The remaining axial-current step is therefore a kinematical low-energy
reduction: it maps the Wilson-completed E3 operator onto the same soft spinor
structure obtained in Sec.~\ref{sec:fieldtheory}, without introducing a new
dynamical transition coefficient.

After the change to the TBTS basis, the external fermions obey
\begin{align}
(i\gamma^\mu D_\mu^+-m)\Psi_i&=0,  \notag \\
(i\gamma^\mu D_\mu^--m)\Psi_f&=0,\qquad m=\frac{\ell}{2\pi\alpha^{\prime}}.
\label{eq:stringDirac}
\end{align}
To analyze the pseudoscalar channel, we define the axial-vector
and pseudoscalar bilinears
\begin{equation}
J_5^\mu=\bar\Psi_f\gamma^\mu\gamma^5U_\gamma\Psi_i,
\qquad
P_5=\bar\Psi_f\gamma^5U_\gamma\Psi_i.
\label{eq:stringJP}
\end{equation}
Using the two Dirac equations together with the derivative of the Wilson
transporter gives
\begin{equation}
\partial_\mu J_5^\mu=2imP_5
+iq\bar\Psi_f\gamma^5\slashed{\mathcal{A}}U_\gamma\Psi_i.
\label{eq:stringaxial}
\end{equation}
This is the Wilson-completed counterpart of the mixed axial identity in
Sec.~\ref{sec:fieldtheory}, now written in the gauge-invariant D7 realization
of the same relative connection $\mathcal A_\mu$. Its sign and the relation to
$\mathcal A_\mu=-\int_\gamma F_{\mu a}dy^a$ are checked explicitly in Appendix~\ref{app:identity}.

For normalizable stationary states in a gauge where the background and $U_\gamma$ are time independent, the spatial integral of the divergence
term is a surface term, while its time component is proportional to the
complete energy difference. At resonance, $E_i=E_f$, one therefore has
\begin{equation}
P_5=-\frac{q}{2m}\,
\bar\Psi_f\gamma^5\slashed{\mathcal A}U_\gamma\Psi_i
\label{eq:stringPsolved}
\end{equation}
inside the stationary transition matrix element. Equation~(\ref{eq:stringPsolved})
is the only identity needed for the leading gauge-dependent amplitude.

\subsection{Low-energy matching of the Wilson-completed E3 operator}

At first order in the interbrane mixing, the Wilson-completed transition
operator contains the pseudoscalar and scalar channels
\begin{equation}
\langle f|\mathcal W_{\rm W}|i\rangle=
-ig\!\int d^3x\,P_5
-im_r\!\int d^3x\,\bar\Psi_fU_\gamma\Psi_i.
\label{eq:stringbareW}
\end{equation}
Inserting Eq.~(\ref{eq:stringPsolved}) at complete-energy resonance gives
\begin{align}
\langle f|\mathcal W_{\rm W}|i\rangle={}&
\frac{iqg}{2m}\!\int d^3x\,
\bar\Psi_f\gamma^5\slashed{\mathcal A}U_\gamma\Psi_i \notag\\
&-im_r\!\int d^3x\,\bar\Psi_fU_\gamma\Psi_i.
\label{eq:stringmasterW}
\end{align}
The plane-wave representation of Eq.~(\ref{eq:stringmasterW}) contains
the Fourier transforms of the complete $x$-dependent products,
\begin{align}
\mathcal{M}_{+\to-}^{(\mathrm{W})}={}&\frac{iqg}{2m}\,
\bar u_-(p^{\prime})\gamma^5
\bigl[\slashed{\mathcal A}U_\gamma\bigr](k)u_+(p)
\notag \\
&-i m_r\,\bar u_-(p^{\prime})\bigl[U_\gamma\bigr](k)u_+(p),
\label{eq:stringmastergeneral}
\end{align}
where the square brackets denote the four-dimensional Fourier transform of
the enclosed product. For the pure four-dimensional vector fluctuation used
in the leading soft matching, one may choose axial gauge $A_a=0$ along
$\gamma$, so that $U_\gamma=1$ and $\mathcal A_\mu$ takes the endpoint-difference
form $A_\mu(p_+)-A_\mu(p_-)$ used in the TBTS description. The first term then gives the
field-dependent contribution fixed by the retained chiral mass operator.
The second is the accompanying
scalar channel without an explicit relative-potential factor. Away from
axial gauge its Wilson transporter is required by endpoint covariance. The
model-dependent E3 form factor discussed in Appendix~\ref{app:softVectorDressing} begins one derivative order higher and
therefore does not alter the coefficient displayed here.

Using $m=\ell/(2\pi\alpha^{\prime})$, the leading pseudoscalar Wilson
contribution in this axial gauge is
\begin{equation}
\mathcal{M}_{+\to-}^{(\mathrm{W},P)}=iqg\frac{\pi\alpha^{\prime}}{\ell}\,
\bar u_-\gamma^5\slashed{\mathcal A}(k)u_+.
\label{eq:leadingWilsonAmplitude}
\end{equation}
This is the leading soft, spinor-level counterpart of
Eq.~(\ref{eq:spinoramp}) within the instanton-generated mass sector. Its
overall strength is inherited from the E3 coefficient through
$g=\mu_{+-}/2$, with the microscopic string data encoded in
Eq.~(\ref{eq:muplusminus}). It is not a complete world-sheet amplitude with
an explicit D7 gauge-field insertion. Independent string operators and
subleading gauge form factors require a dedicated calculation.

\paragraph*{Hamiltonian correspondence.}

Equation~(\ref{eq:stringmastergeneral}) reproduces the spinor structures
of Eq.~(\ref{eq:spinoramp}). Applying the nonrelativistic
reduction derived in Sec.~\ref{sec:fieldtheory}, with
\begin{equation}
\hat{\bm\mu}
=\frac{q}{2m}\bm\sigma
=\frac{\pi\alpha^{\prime}q}{\ell}\bm\sigma,
\label{eq:stringMagneticMoment}
\end{equation}
the leading spin-dependent term in the resonant matrix element is
represented by
\begin{equation}
H_{\mathrm{cm}}^{\mathrm{str}}=
\begin{pmatrix}
0 & -ig\hat{\bm\mu}\cdot\bm{\mathcal{A}} \\
ig\hat{\bm\mu}\cdot\bm{\mathcal{A}} & 0
\end{pmatrix}.
\label{eq:stringSwappingHamiltonian}
\end{equation}
This reproduces the TBTS spin structure and its coefficient within the
on-shell nonrelativistic matching. As in Sec.~\ref{sec:fieldtheory}, using
this matrix as an autonomous two-state Hamiltonian requires a specified
projection onto states.

The single-E construction also retains the field-independent scalar companion,
\begin{equation}
H_{\mathrm{c}}^{\mathrm{str}}=
\begin{pmatrix}
0 & i m_r \\
-i m_r & 0
\end{pmatrix},
\qquad m_r=-g,
\label{eq:stringScalarHamiltonian}
\end{equation}
with the formal coefficient ratio
\begin{equation}
\frac{|m_r|}{|g|\,|q|\,|\bm{\mathcal A}|/(2m)}
=\frac{2m}{|q|\,|\bm{\mathcal A}|}.
\label{eq:stringHierarchy}
\end{equation}
This is not a hierarchy of physical transition amplitudes, and no
magnetic-dominated regime is established by this matching. The states and
the validity of their nonrelativistic approximation must also be specified,
as discussed after Eq.~(\ref{eq:rhosingleE}). Independently of this issue,
the ordinary massless D7 zero mode cancels from the relative connection.
The following subsection determines the resulting restriction on the range
of the relative interaction.

\subsection{Structural obstruction and range of the relative interaction}

\label{sec:modes}

The preceding analysis establishes that a common D7 Abelian gauge field
naturally enters the mixed $D3_+$--$D3_-$ transition operator through the
relative connection $\mathcal{A}_\mu$, thereby reproducing the relative
gauge structure that, in the two-sheeted description, is associated with
the difference of the electromagnetic potentials on the two sheets.
This is a local operator statement, however, and does not by itself imply
the existence of a massless four-dimensional relative photon, nor therefore
of a genuinely long-range electromagnetic interaction.

The relevant question is thus whether the minimal connected D7 sector
contains a massless Abelian mode that couples to $\mathcal{A}_\mu$, or
whether the modes capable of distinguishing the two endpoint loci
$p_+$ and $p_-$ are necessarily massive. The latter possibility would
preserve the local transition operator while limiting the range of its
four-dimensional gauge interaction. This question can be decided directly
from the internal mode decomposition.

The common D7 contains one higher-dimensional gauge field. In axial gauge,
expand its four-dimensional components in internal eigenmodes,
\begin{equation}
A_\mu(x,y)=\sum_n f_n(y)a_\mu^{(n)}(x).
\label{eq:D7modeExpansion}
\end{equation}
Then
\begin{equation}
\mathcal{A}_\mu =
\sum_n\left[f_n(p_+)-f_n(p_-)\right]a_\mu^{(n)}.
\label{eq:relativeModeExpansion}
\end{equation}

The magnitude of the relative connection is controlled not only by the
four-dimensional mass of the contributing mode but also by its variation
between the two endpoint loci. For a smooth mode whose internal variation
scale is large compared with the separation $d$, the symmetric configuration
of Eq.~(\ref{eq:pmp}) gives
\begin{equation}
f_n(p_+)-f_n(p_-)
=
d\,\partial_{x^4}f_n(0)
+\mathcal{O}\!\left(d^3\partial_{x^4}^3f_n\right).
\label{eq:relativeProfileGradient}
\end{equation}
Thus a light mode is not by itself sufficient: its internal profile must also
vary appreciably between $p_+$ and $p_-$ for it to generate a sizeable
relative connection.

For the ordinary unbroken D7 Abelian vector on a compact connected internal
divisor, the massless four-dimensional mode has an internal profile $f_0$ in
the kernel of the scalar Laplacian. Harmonic functions on a compact
connected space are constant, so $f_0(y)=\mathrm{const}$ and
Eq.~(\ref{eq:relativeModeExpansion}) gives
$\mathcal{A}_\mu[f_0]=0$ identically. This is a
structural obstruction, not merely an additional phenomenological spectral
condition. The same result is manifest in the gauge-invariant form Eq.~(\ref%
{eq:relativeConnection}): an internally constant vector mode has $F_{\mu a}=0$ and
therefore makes no contribution to $\mathcal{A}_\mu$, independently
of the axial-gauge choice.

Equivalently, the Wilson-completed mixed operator is neutral under the
four-dimensional D7 $U(1)$ associated with constant internal gauge
transformations, so only internally varying gauge data can enter the
relative connection.

Consequently, every mode satisfying $f_n(p_+)\neq f_n(p_-)$ has non-zero
internal eigenvalue and is a Kaluza--Klein (KK) excitation. In a generic compact
geometry its mass is set by the compactification scale, and the
corresponding relative interaction has a range set by the inverse mass of
the contributing mode. This does not require all non-uniform modes to share
one universal ``KK scale'': warping or special spectral geometry may leave
the lowest non-uniform eigenmode parametrically lighter than the rest of the
tower. An exactly massless relative mode is nevertheless absent for a single
compact connected D7. A phenomenologically long-range effect therefore
requires additional compactification structure producing an anomalously
light non-uniform D7 gauge mode. In a geometry symmetric under $%
p_+\leftrightarrow p_-$, even modes cancel and odd modes contribute.

A connected D7 therefore provides the Wilson transporter but not a massless
relative photon. Disconnected D7 components could support independent
Abelian zero modes, but then the mixed bilinear requires additional link
degrees of freedom to restore gauge covariance. This is a limitation of the
minimal construction, not a no-go theorem for more elaborate D7 sectors.

This conclusion excludes only the ordinary constant zero mode of a single
compact connected D7 as the source of the relative connection. It does not
preclude an effectively long-ranged relative gauge field arising from an
anomalously light non-uniform mode or from a more elaborate
localization/compactification mechanism. For example, a quasi-localization
mechanism of the Dvali--Gabadadze--Shifman type \cite{DGS} could in principle
modify the infrared gauge spectrum, although no such mechanism is assumed
in the present construction.

The local transition operator itself requires only a non-zero relative
connection and does not assume that it is long-ranged. Producing two 
approximately autonomous four-dimensional ``sheet photons''
would require additional compactification input, including the spectrum
and possible four-dimensional Stueckelberg mass of the relevant D7 Abelian
combinations. None of these additional requirements is assumed in deriving the Wilson
contribution Eq.~(\ref{eq:leadingWilsonAmplitude}).
It is important to distinguish this limitation from a second, independent 
requirement concerning electromagnetic sequestering. In the domain-wall 
realization of Ref.~\cite{PRD2010}, the gauge fields are localized separately 
on the two branes and the coupling of matter to the gauge field localized on 
the opposite brane is exponentially suppressed. In the minimal D3$_+$--D7--D3$_-$ 
configuration considered here, by contrast, the two D3--D7 fermions carry the 
same physical charge under the common D7 Abelian gauge field. The present 
construction therefore does not by itself realize electromagnetic invisibility 
of one fermion sector with respect to the other. In particular, although an 
internally constant D7 zero mode cancels from the relative connection 
$\mathcal{A}_\mu=A_\mu(p_+)-A_\mu(p_-)$ and hence does not contribute to the 
swapping operator, it still couples diagonally to both charged sectors. Recovering 
mutually sequestered visible and hidden sectors therefore requires additional 
compactification or localization structure suppressing the coupling of each sector 
to the photon mode observed by the other, while retaining the relative gauge 
structure needed for the transition operator. This requirement is distinct 
from the need for a sufficiently light non-uniform D7 mode discussed above.

\section{Status, scope, and roadmap to a compact realization}

\label{sec:roadmap}

The construction developed in the present work is local. It identifies
the microscopic ingredients needed to reproduce the two-sheeted
transition operator and establishes the corresponding four-dimensional
matching under explicit assumptions on the instanton spectrum and
charged disk couplings. A complete string realization, however,
requires more. These ingredients must be compatible within a common
local D3$_+$--D7--D3$_-$/E3 configuration. This configuration must then
admit an embedding into a globally consistent compactification in which
the required brane sectors, instanton zero modes, gauge selection rules,
and light four-dimensional fields survive simultaneously.

The purpose of this section is therefore twofold. First, we distinguish
the results established in this work from the local and global
conditions that remain to be met. Second, we formulate a roadmap
toward a compact model in which the mixing coefficients and the
relative gauge interaction can be calculated from microscopic data.
Table~\ref{tab:statusRoadmap} organizes this program into three stages:
the local microscopic mechanism, its matching to the TBTS interaction,
and the required gauge sector and compactification. For each target,
it lists the key requirement, the present status, and the remaining task.

\begin{table*}[!tp]
\caption{Roadmap from the local D3$_+$--D7--D3$_-$/E3 mechanism
to a complete compact realization. The status column distinguishes
results established under the stated assumptions from the remaining
local and global construction tasks. The separate orbifold example
provides explicit checks of several charged ingredients.}
\label{tab:statusRoadmap}
\centering
{\footnotesize
\begin{tabular}{
p{0.25\textwidth}
p{0.31\textwidth}
p{0.36\textwidth}}
\toprule
\textbf{Objective}
&
\textbf{Key requirement / criterion}
&
\textbf{Status and next step}
\\
\midrule

\multicolumn{3}{l}{
\textit{Stage I: Establish the local microscopic mechanism}}
\\[2pt]

Realize two D3--D7 fermion sectors with equal physical masses
&
Four-ND D3--D7 sectors with equal DD lengths and the coupling
$\widetilde Q_s(\Phi_7^3-\Phi_s^3)Q_s$, with
$z_{3,+}=-z_{3,-}$
&
\textbf{Established for the smooth array at tree level.}
$|m_\pm|=m=\ell/(2\pi\alpha')$, with $m_+=-m_-$ in the
chosen holomorphic basis. A common local realization must
preserve these massive sectors and their separated endpoints
under the required projection.
\\[5pt]

Forbid direct perturbative mass mixing between the two D3 sectors
&
Chan--Paton non-closure of
$Q_+\widetilde Q_-$ and $Q_-\widetilde Q_+$
&
\textbf{Established.}
The stated minimal matter content obeys an exact perturbative
selection rule forbidding both bilinears.
\\[5pt]

Generate the off-diagonal chiral mass
$Q_+\widetilde Q_-$ non-perturbatively
&
Compatible local projection and RR couplings, together with
divisor and bundle data yielding the required charged
multiplicities, neutral $d^4x\,d^2\theta$ measure, and
non-vanishing full charged disk coefficients
&
\textbf{Derived under the stated instanton assumptions.}
The orbifold example checks E3--D3 modes, E3--D7
representations, allowed disk channels and RR charge.
These ingredients must be made compatible with the massive
separated-endpoint configuration. The full measure and disk
amplitudes must then be determined using the required
local and compact data.
\\

\midrule
\multicolumn{3}{l}{
\textit{Stage II: Match the string operator to the TBTS interaction}}
\\[2pt]

Identify the instanton-induced mass with the TBTS scalar and
pseudoscalar mixing coefficients
&
Conversion from the natural string basis to the
equal-positive-mass Dirac basis
&
\textbf{Established algebraically.}
For the assumed single-$E$ chiral mass,
$g=\mu_{+-}/2$ and $m_r=-g$ after the residual vector rephasing.
\\[5pt]

Make the mixed D3$_+$--D3$_-$ operator gauge covariant and identify
the relative gauge connection
&
Wilson completion of the bilinear and endpoint gauge covariance
&
\textbf{Established kinematically.}
For a chosen reference path, the Wilson-completed D7 connection
reproduces the relative gauge connection entering the TBTS operator.
\\[5pt]

Recover the spin--magnetic-moment swapping interaction
$H_{\mathrm{cm}}^{\mathrm{str}}
 \propto g\,\hat{\bm\mu}\!\cdot\!\bm{\mathcal A}$
&
Single-$E$ chiral mass, positive-mass conversion, Wilson completion,
and the mixed axial identity for the leading
$O(k^0\delta A)$ contribution of this mass operator
&
\textbf{Established for stationary resonant matrix elements
at leading order under the stated inputs.}
The scope of the soft-momentum expansion is specified in Appendix~\ref{app:softVectorDressing}. An explicit D7
gauge-field insertion is needed to determine subleading
dressing and possible independent string contributions.
\\[5pt]

Determine whether the phenomenology is dominated by magnetic
swapping rather than by the scalar companion
&
The single-$E$ relation $m_r=-g$, the gauge background and
specified external states within a controlled low-energy regime
&
\textbf{Not inferred from coefficient matching.}
The scalar companion is retained. The two transition
amplitudes must be compared for specified spatial and spin
states, with the nonrelativistic restriction discussed after
Eq.~(\ref{eq:rhosingleE}) taken into account.
\\

\midrule
\multicolumn{3}{l}{
\textit{Stage III: Realize the required gauge sector and global compactification}}
\\[2pt]

Obtain an effectively long-ranged relative gauge interaction
&
A sufficiently light D7 gauge mode with a non-uniform internal
profile, $f_n(p_+)-f_n(p_-)\neq0$, and a four-dimensional
mass compatible with the desired interaction range
&
\textbf{Not supplied by the constant Abelian zero mode.}
For the compact connected D7 considered here,
$f_0=\mathrm{const}$ gives $\mathcal A_\mu[f_0]=0$.
The next step is to identify a compactification or localization
mechanism that supplies a sufficiently light non-uniform mode
and to calculate its endpoint couplings.
\\[5pt]

Electromagnetically sequester the two fermion sectors while
retaining the relative gauge structure required for swapping
&
Compactification or localization structure suppressing the
coupling of each sector to the photon mode observed by the
other, while preserving the relative connection
&
\textbf{Not realized by the minimal common-D7 array.}
The two D3--D7 fermions have the same physical D7 charge.
Cancellation of the constant mode from $\mathcal A_\mu$
does not remove its diagonal couplings. Sequestering must
suppress each sector's coupling to the photon mode observed
by the other while preserving the required relative interaction.
\\[5pt]

Embed all ingredients simultaneously in a globally consistent
compact string model
&
Compatible brane and orientifold data, supersymmetry conditions
on fluxes, Freed--Witten quantization, RR tadpole cancellation,
K-theory and four-dimensional anomaly constraints, and survival
of the required instanton and gauge sectors
&
\textbf{A compact realization is not constructed here.}
A compact embedding must satisfy these conditions and
retain the local mechanism. Any additional sectors must
be included when checking the instanton zero modes
and their saturation.
\\

\bottomrule
\end{tabular}
}
\end{table*}

The first construction task concerns the local instanton sector.
The projected charged representations and the compensating gauge
charge of the instanton weight must be compatible with the massive
D3--D7 sectors and their separated endpoints. The orbifold example
in Appendix~\ref{app:orbifold} explicitly checks several of these
charged ingredients, including the relation between the E3--D3
spectrum and the gauge transformation of the instanton weight.
As explained in Appendix~\ref{app:orientifoldscope}, it is a separate
setting and does not provide the full configuration used in the
operator calculation. The next step is therefore to construct a local
projection that preserves the two independent massive D3--D7 sectors
and gives compatible charged representations and RR couplings.
Determining the complete instanton measure requires further data.
In particular, the E3--D7 zero-mode multiplicities depend on the
intersection curve and its bundles or boundary conditions, while
the neutral modes depend on the wrapped divisor, its projection,
and fluxes. These data must yield the assumed measure and
non-vanishing charged disk amplitudes. Local compatibility is
therefore necessary, but does not by itself settle the full
zero-mode problem.

A compact realization must also satisfy the usual orientifold, flux,
Freed--Witten, tadpole, K-theory, and four-dimensional anomaly
constraints
\cite{BianchiCollinucciMartucci,GrimmFluxedInstantons,FreedWitten,UrangaKTheory,IbanezRabadanUrangaU1}.
For a reader interested primarily in the four-dimensional mechanism,
these conditions govern the compatibility of the brane and flux
content with supersymmetry, charge quantization, charge cancellation,
and the absence of anomalies. The task is to satisfy them while
preserving the local sectors needed for the mixing. Any additional
branes or matter introduced in the compact model must also be included
when checking the instanton zero modes and their saturation.

On the gauge side, the ordinary unbroken Abelian vector on a compact
connected D7 has a constant zero-mode profile, which cancels from
the relative connection $\mathcal A_\mu$.
A long-range relative interaction therefore requires a sufficiently
light non-uniform D7 mode with different values at the two endpoints.
The compact gauge spectrum and its internal profiles must be calculated
to determine whether such a mode exists and what interaction range it
provides. A hidden-sector interpretation requires a further ingredient.
The two local D3--D7 fermions have the same physical D7 charge, so the
cancellation of the constant mode from $\mathcal A_\mu$ does not suppress
their ordinary coupling to the common photon. Electromagnetic
sequestering must suppress the coupling of each sector to the photon
mode observed by the other without removing the relative interaction.

The relative strength of magnetic and scalar transitions can then be
assessed for the chosen gauge background and external fermion states.
The single-$E$ relation $m_r=-g$ fixes the coefficients of the two
operators, but their transition amplitudes also depend on spatial and
spin overlaps. Both contributions must be retained, and any proposed
magnetic-dominated regime must be checked within the domain of validity
of the low-energy expansion. In particular, for homogeneous spatial
backgrounds and constant mixing, the condition
$|q|\,|\bm{\mathcal A}|/(2m)\gg1$ is incompatible with both coupled
plane-wave states being nonrelativistic at the same canonical momentum,
as explained after Eq.~(\ref{eq:rhosingleE}).
Determining subleading E3 gauge dressing and possible independent
string operators also requires an explicit calculation with a D7
gauge-field insertion or E3--D7 overlap wavefunctions, as discussed
in Appendix~\ref{app:softVectorDressing}.

Taken together, these steps identify the two principal microscopic
targets of a complete realization. An explicit instanton calculation
would determine the holomorphic mixing coefficient
$\mu_{+-}^{\rm hol}$ of Eq.~(\ref{eq:muplusminus}) from the instanton
action, one-loop factor, and charged disk amplitudes. Within the
single-$E$ effective theory and normalization assumptions retained
here, canonical normalization would then give the physical coefficient
$\mu_{+-}$ and hence $g=\mu_{+-}/2$ and $m_r=-g$ through
Eq.~(\ref{eq:gmfrommu}). The compact D7 spectrum and its endpoint
profiles would determine the relative gauge coupling and its
interaction range. A successful implementation of this roadmap would
thus turn the conditional operator matching into a model with
calculable mixing and gauge parameters, whose transition predictions
could be compared with low-energy searches in the corresponding
physical regime.

\FloatBarrier

\section{Conclusions}
\label{sec:conclusion}

We have investigated the conditions under which the characteristic 
fermion-mixing operator of the infrared two-sheeted description of a
two-brane Universe can emerge from a local Type IIB
D3$_+$--D7--D3$_-$/E3 configuration. The two fermion
sectors are distinguished by their D3 endpoints while
sharing a common D7 gauge brane. For the minimal matter
content considered here, direct mixing is perturbatively
forbidden. Under the required instanton zero-mode and
disk-amplitude conditions, however, an E3 instanton
generates a chiral off-diagonal mass term. After conversion
to equal positive Dirac masses, this term gives both
pseudoscalar and scalar mixing contributions, with
$g=\mu_{+-}/2$ and $m_r=-g$ in the single-$E$ case.
For distinct D7 endpoint loci, Wilson transport provides
the gauge-invariant relative connection entering the
two-sheeted description. The mixed axial identity then
relates the mass operator to the leading
spin--magnetic-moment transition contribution for
stationary resonant matrix elements, under the stated
low-energy assumptions. The main result is thus a conditional effective-field-theory 
operator matching: provided the specified microscopic conditions 
are satisfied, a non-perturbative D-brane mechanism 
reproduces the local TBTS operator structure.

An important aspect of the analysis is the identification
of the instanton ingredients needed for this mechanism.
The smooth brane array does not by itself supply the
required chiral E3--D3 spectrum or the compensating
gauge charge of the instanton weight.
The complementary orbifold example illustrates how
these charged ingredients can arise and explicitly
checks the E3--D3 modes, the local E3--D7 representations,
the allowed disk channels and the compensating RR charge.
It does not, however, constitute a complete
realization of the massive separated-endpoint system. In particular, 
the simultaneous realization of the chiral E3--D3 spectrum, the compensating RR charge, the neutral 
instanton measure and two independent massive D3--D7 sectors 
in a single consistent local configuration has not been 
demonstrated. The full instanton measure and non-vanishing amplitudes
must also be established using the relevant divisor,
bundle and compactification data, within a background
satisfying the global consistency conditions.

The phenomenological interpretation requires further
care. In particular, the constant Abelian mode of a
compact connected D7 cancels from the relative
connection. A long-range relative interaction therefore
requires a sufficiently light non-uniform gauge mode
whose profile takes different values at the two endpoints.
Moreover, the minimal common-D7 construction assigns
the same physical D7 charge to both fermions, so an
electromagnetically hidden sector requires additional
sequestering. The relative strength of magnetic and
scalar transitions also depends on the gauge background
and the external states. Although $m_r=-g$ fixes the
relation between their coefficients, it does not by
itself imply that magnetic transitions dominate.

Finally, the analysis provides a roadmap for future work,
summarized in Sec.~\ref{sec:roadmap}. The first task is to
construct a local D3$_+$--D7--D3$_-$/E3 configuration
with compatible matter sectors, instanton representations
and RR couplings. The full zero-mode measure and non-vanishing charged disk amplitudes
must be established for this same configuration, and its compact
embedding must satisfy the global consistency conditions. Such a realization 
would allow the holomorphic mixing coefficient $\mu_{+-}^{\rm hol}$ to be calculated
from the instanton action, the one-loop factor and the
charged disk amplitudes. Under the normalization assumptions
retained here, canonical normalization would give the
physical coefficient $\mu_{+-}$ and hence
$g=\mu_{+-}/2$ and $m_r=-g$ within the single-$E$
effective theory. In parallel, the D7 gauge spectrum and
its endpoint profiles would determine the relative gauge
coupling and its interaction range. These steps would
turn the conditional operator matching established here
into a quantitative string realization of two-sheeted
fermion mixing. Its transition predictions could then
be compared with low-energy experimental constraints
in the corresponding physical regime.

\section*{Generative AI Assistance and Author Verification}

The author made substantial use of OpenAI ChatGPT (GPT-5.6 Sol and Astra 6) and
Anthropic Claude (Opus 5 and Fable 5) as interactive research assistants.
Their use included mathematical and symbolic reasoning, verification of
intermediate derivations, exploration of alternative theoretical
constructions, literature identification, consistency
auditing, and assistance with drafting, restructuring, and revision.
The process remained author-directed: the author formulated the physical
questions, selected or rejected assumptions and theoretical directions,
and determined which arguments were retained. Where useful, the AI models 
were queried independently or employed in multi-agent 
workflows, with their analyses compared to cross-check derivations, 
expose hidden assumptions, and assess alternative reasoning.

The TBTS formalism, the physical framework, the D3--D7--D3 strategy,
the interpretation of the results, and the scientific objectives and
conclusions were conceived and determined by the author. All figures were
conceived, designed, and produced by the author. All retained AI-assisted
material was critically reviewed for mathematical and physical consistency,
with selected derivations and calculations independently re-derived or
recomputed when needed. Proposals that could not be sufficiently justified
or verified were discarded or reformulated as assumptions, sufficient
conditions, or open problems. The AI models were not treated as authoritative
sources, and all references, scientific claims, interpretations, and
conclusions were accepted only after author review. The author assumes full
and sole responsibility for the accuracy, originality, interpretation, and
conclusions of the work.

\section*{Acknowledgments}
The author acknowledges Micha\"{e}l Lobet for valuable discussions on the use 
of artificial intelligence in research, including research-use protocols and 
the appropriate disclosure of AI-assisted contributions in scientific work.

\section*{Funding} 
The author declares that no funds, grants, or other support
were received during the preparation of this manuscript.

\section*{Data Availability Statement}
This manuscript has no associated data.

\section*{Code Availability Statement}
This manuscript has no associated code/software.

\appendix

\section{E3/D3/D7 ND counting and Chan--Paton bookkeeping}

\label{app:CFT}

This appendix records the explicit boundary-condition and picture-number
checks underlying Secs.~\ref{sec:E3candidate} and~\ref{sec:chargedmodes},
without repeating the main-text Chan--Paton algebra.

The ordinary $37$ sectors have four mixed ND directions, so their
supersymmetric ground-state spectrum is the standard D3--D7 hypermultiplet
\cite{Polchinski:1998rq,Bachas:1998rg}.
The DD separation in $z_3$ shifts its physical mass to $m=\ell/(2\pi\alpha^{%
\prime })$ without changing that oscillator structure. For the local $E3$
of Eq.~(\ref{eq:E3divisor}), the E3--D3 strings are mixed in the four
spacetime directions and the four real directions of $z_1,z_3$, whereas
E3--D7 strings are mixed in spacetime, $z_2$ and $z_3$. Both sectors
therefore have eight ND directions, giving the open-string zero-point energies
$a_{\mathrm{NS}}=+1/2$ and $a_{\mathrm{R}}=0$. The NS and R sectors
produce spacetime bosons and fermions, respectively. Hence no massless
bosonic NS zero mode is present, while R-sector ground states can furnish the
charged Grassmann zero modes used in the instanton measure
\cite{BCKW,CveticRichterWeigand}.

For E3--D7 there are two common NN real directions along $C_{E7}$.
Consequently the actual four-dimensional multiplicity is determined by
zero-eigenvalue wavefunctions on the compact intersection curve after the
GSO physical-state projection, orientifold/orbifold identifications and the
internal gauge-bundle projection. The local ND count does not replace this
global zero-mode (cohomology) calculation, nor does $a_{\mathrm{R}}=0$
determine the neutral $O(1)$ measure: possible deformation and internal
Wilson-line zero modes, including their flux-modified versions, must still be
checked on the compact divisor
\cite{BCKW,BianchiCollinucciMartucci,GrimmFluxedInstantons}.
By contrast, E3--D3 strings have no common NN direction, so their zero
modes are localized at the D3 position and are insensitive to boundary
conditions at infinity \cite{IbanezUrangaInstanton}. Each orientation
contributes one R-sector state. Which orientations survive is decided by the
local projection (see Appendix~\ref{app:chirality}).

For completeness, we now translate the endpoint notation of the main text into 
standard Chan--Paton matrices. Let $e_{ab}$ denote the matrix unit associated 
with an oriented open string from boundary $a$ to boundary $b$. It is a bookkeeping 
matrix, not a physical field. Then the perturbative cross-bilinears fail to close,
\begin{equation}
\mathrm{Tr}(e_{+7}e_{7-})=\mathrm{Tr}(e_{+-})=0,\qquad
\mathrm{Tr}(e_{-7}e_{7+})=0,
\label{eq:CPtwo}
\end{equation}
whereas the two three-point instanton disks close on the E3 boundary,
\begin{equation}
e_{E+}e_{+7}e_{7E}=e_{EE},\qquad
e_{E7}e_{7-}e_{-E}=e_{EE}.
\label{eq:E3CPclosure}
\end{equation}
Here $E$ is only the label of the E3 boundary. For an $O(1)$ instanton there is 
no four-dimensional $U(1)_E$ gauge field. The orientations merely encode charges under the space-filling D3 and D7 branes.

Finally, the total picture number of each disk in Eq.~(\ref{eq:E3disks}) is $%
-2$: the two instanton zero modes are Ramond vertices in the $-1/2$ picture
and the matter insertion is the NS scalar in the $-1$ picture. ``Picture'' is
only the standard world-sheet superghost bookkeeping convention. It is not a
new four-dimensional quantum number. This is the standard charged-zero-mode
construction of a holomorphic superpotential
\cite{CveticRichterWeigand,IbanezUrangaInstanton,BCKW}. The fermion bilinear follows
after component expansion.

\section{Chirality of the E3--D3 zero modes and the instanton charge}

\label{app:chirality}

This appendix provides an illustration of how the charged
ingredients required by the proposed mass operator can arise.
It does not construct a complete realization of the massive
separated-endpoint configuration studied in the main text.
We first explain why the smooth array does not supply the
required chiral spectrum and instanton charge
(see Appendix~\ref{app:smooth}).
We then consider a separate orbifold example in which a chiral
E3--D3 spectrum and a compensating RR transformation can be
derived explicitly (see Appendix~\ref{app:orbifold}).
This example illustrates these ingredients without establishing
the complete minimal charged zero-mode spectrum assumed in
the main text.
Its scope, including the distinction between local E3--D7
representations and zero-mode multiplicities and the role
of the orientifold projection, is specified in
Appendix~\ref{app:orientifoldscope}.

\subsection{D3-branes at smooth points of the instanton divisor}

\label{app:smooth}

For D3$_s$ at a smooth point of $\Sigma_E$, the E3--D3$_s$ string has eight
ND directions and one DD complex plane, $z_2$ (see Appendix~\ref{app:CFT}). Each
orientation provides one R-sector fermionic zero mode, and the local array
contains no projection that could remove one of them. Besides the modes
$\lambda_+$ and $\lambda_-$ of Eq.~(\ref{eq:minzeromodes}), the instanton
therefore carries
\begin{equation}
\lambda'_+:\ D3_+\to E3,\qquad \lambda'_-:\ E3\to D3_-,
\label{eq:smoothpartners}
\end{equation}
with charges $+1$ under $U(1)_{3+}$ and $-1$ under $U(1)_{3-}$,
respectively. Each pair couples to the relative position of D3$_s$ and the
instanton in the DD plane,
\begin{equation}
S_{\mathrm{ch}}\supset\kappa_+\,Z_+\,\lambda_+\lambda'_+
+\kappa_-\,Z_-\,\lambda'_-\lambda_- ,
\label{eq:Ganorcoupling}
\end{equation}
Here $Z_s$ is the local complex displacement of D3$_s$
from the E3 in the transverse $z_2$ plane. The local
coupling has the form derived in
Ref.~\cite[Sec.~2.1]{Ganor}, where two charged fermionic
variables couple to the complex separation through
$z\psi_1\psi_2$. The coefficients $\kappa_s$ absorb the
normalization of the displacement and of the fermionic
variables and are non-zero for this local coupling.
Integrating over an isolated pair therefore gives a
factor proportional to $Z_s$, which vanishes when the
D3-brane reaches the instanton divisor. In the
configuration of Fig.~\ref{fig:localgeometry},
$Z_\pm=0$, so these displacement couplings do not lift
either pair.

The charged zero modes $\{\lambda_\pm,\lambda'_\pm,\eta,\widetilde\eta\}$
then have vanishing total charge under
$U(1)_{3+}\times U(1)_{3-}\times U(1)_7$, and every coupling in
$S_{\mathrm{ch}}$ closes on the E3 boundary. Every operator obtained by
saturating the Grassmann measure is therefore neutral under
$U(1)_{3+}\times U(1)_{3-}$, and $Q_+\widetilde Q_-$, of charge $(+1,-1)$, is
not generated. A saturation containing $Q_+\widetilde Q_-$ must therefore
include additional matter insertions carrying the opposite
D3 charges. A schematic example of the required charge
compensation is
\begin{equation}
\big(\lambda_+Q_+\eta\big)\big(\widetilde\eta\,\widetilde Q_-\lambda_-\big)
\big(\lambda'_-X_{-+}\lambda'_+\big)\ \longrightarrow\
Q_+\widetilde Q_-X_{-+},
\label{eq:smoothneutral}
\end{equation}
Here $X_{-+}$ denotes a D3$_-\to$D3$_+$ matter insertion
with charges $(-1,+1)$ under
$U(1)_{3+}\times U(1)_{3-}$. The arrow indicates the
Grassmann saturation if the displayed couplings are
present, with their coefficients suppressed. It does
not assert a non-vanishing disk amplitude for the
separated configuration. No expectation value for
$X_{-+}$ is assumed. The obstruction to the isolated
bilinear follows from charge neutrality independently
of this illustrative saturation.
This neutrality is consistent with the absence of an axionic shift of the
instanton action under the D3 trace $U(1)$ at a smooth point. In orbifold
models, the fields shifted by the $U(1)$s of D3-branes are twisted-sector RR
fields localized at the singularities
\cite{IbanezRabadanUrangaU1,IbanezUrangaInstanton}, and no such field is
available at a smooth point.

\subsection{An illustrative orbifold calculation of the charged sector}

\label{app:orbifold}

We consider fractional D3-branes at a common
$\mathbb{C}^3/\mathbb{Z}_N$ singularity to examine explicitly
the relation between the orbifold projection, the charged
open-string representations and the axionic transformation
of the E3 action. The calculation identifies the representations
entering the two charged disk channels and compares the
instanton charge obtained from the RR couplings with the
total E3--D3 zero-mode charge. A $\mathbb{Z}_3$ specialization
makes these relations explicit.

We use the conventions of Refs.~\cite{AIQU,IbanezUrangaInstanton}. The
generator $\theta$ of $\mathbb{Z}_N$ acts on the fermions of the parent
$\mathcal{N}=4$ theory with phases $e^{2\pi ia_r/N}$, $r=1,2,3$, with
$a_4=0$ for supersymmetry, and on $z_r$ with phase $e^{-2\pi ia_r/N}$. We
take $N$ odd and choose integer representatives with $a_1+a_2+a_3=0$ and all
$a_r$ even. This is always possible for odd $N$, since the odd entries come
in pairs that can be shifted by $\pm N$, for instance $(1,1,-2)\to(-2,-2,4)$
for $\mathbb{Z}_3$. Then $a_r/2$ is an integer, Chan--Paton matrices of the
form (\ref{eq:orbCP}), with $\gamma_\theta^N=\mathbf 1$, can be used for all
branes, and the even-$a_r$ formulas of
Refs.~\cite{AIQU,IbanezUrangaInstanton} apply throughout. We also assume
$a_r\not\equiv0$ (mod $N$) for $r=1,2,3$, so that the orbifold acts on all
three planes. The Chan--Paton matrices are
\begin{equation}
\gamma_{\theta,3}=\mathrm{diag}\big(\mathbf 1_{n_0},\alpha\mathbf 1_{n_1},
\ldots,\alpha^{N-1}\mathbf 1_{n_{N-1}}\big),\qquad \alpha=e^{2\pi i/N},
\label{eq:orbCP}
\end{equation}
and similarly with multiplicities $u_j$ for the D7 transverse to $z_3$ and
$v_j$ for the E3 transverse to $z_2$.
Here, an orbifold \emph{node} $i\in\mathbb{Z}_N$
denotes the Chan--Paton sector associated with
the eigenvalue $\alpha^i$ of the orbifold generator,
equivalently a vertex of the corresponding quiver.
The multiplicities $n_i$, $u_i$, and $v_i$
specify the dimensions of the corresponding
D3, D7, and E3 Chan--Paton spaces, respectively.
All node indices are understood modulo $N$.
The D7 and E3 divisors have the orientations
specified in Eqs.~(\ref{eq:D7divisor})
and~(\ref{eq:E3divisor}), respectively.
Their common NN complex plane is $z_1$,
corresponding to the direction of $C_{E7}$.
We denote by $(x_i,\bar y_j)$ an
oriented string from node $i$ of brane $x$ to node $j$ of brane $y$,
transforming in the fundamental of the first and the antifundamental of the
second, in agreement with Table~\ref{tab:zeromodecharges} and with the
matrix units $e_{ab}$ of Appendix~\ref{app:CFT}.

For completeness, the orbifold projection on a Chan--Paton wavefunction
$\Lambda_{xy}$ in each relevant ground-state sector is
\begin{equation}
\Lambda_{xy}
=\alpha^{-a_r/2}\gamma_{\theta,x}\Lambda_{xy}
\gamma_{\theta,y}^{-1}.
\label{eq:orbGroundProjection}
\end{equation}
The phase $\alpha^{-a_r/2}$ is the orbifold eigenvalue
of the relevant ground state in the conventions used
here. For the E3--D3 fermions, $r=2$ labels the DD
complex plane. For the E3--D7 fermions, $r=1$ labels
the common NN complex plane. These assignments give
the even-twist spectra of
Ref.~\cite[Eqs.~(2.2) and~(2.4)]{IbanezUrangaInstanton}.
For D3--D7 matter, $r=3$, consistently with the
projection in Ref.~\cite[Sec.~2.2]{AIQU}.

The Chan--Paton part of the orbifold action on the
$(i,j)$ block of $\Lambda_{xy}$ is given by
\[
\left(\gamma_{\theta,x}\Lambda_{xy}
\gamma_{\theta,y}^{-1}\right)_{ij}
=\alpha^{i-j}(\Lambda_{xy})_{ij}.
\]
Equation~(\ref{eq:orbGroundProjection}) therefore
retains this block only if
$i-j-a_r/2\equiv0\pmod N$.
Applying this condition to both orientations gives
\begin{align}
\text{E3--D3:}&\quad \sum_i\big[(n_i,\bar v_{i-a_2/2})
+(v_i,\bar n_{i-a_2/2})\big], \notag\\
\text{E3--D7:}&\quad \sum_i\big[(u_i,\bar v_{i-a_1/2})
+(v_i,\bar u_{i-a_1/2})\big], \label{eq:orbspectra}\\
\text{D3--D7:}&\quad \sum_i\big[(n_i,\bar u_{i-a_3/2})
+(u_i,\bar n_{i-a_3/2})\big]. \notag
\end{align}
The three lines reproduce, respectively,
Eqs.~(2.2), (2.4) and~(A.5) of
Ref.~\cite{IbanezUrangaInstanton} for the divisor
orientations chosen here. The first line describes
localized E3--D3 modes. The second describes E3--D7
fermionic fields on the intersection curve, whose
zero-mode multiplicities depend on boundary conditions
or compactification data, as discussed in
Ref.~\cite[Sec.~2.3]{IbanezUrangaInstanton}.
The third line gives the D3--D7 matter chiral multiplets.

We choose $v_0=1$ and $v_{j\neq0}=0$. This Chan--Paton representation is compatible
with a candidate $O(1)$ instanton sector under
the conventional orientifold action conjugating
Chan--Paton phases
\cite{IbanezUrangaInstanton}.
However, this choice does not itself implement
the orientifold projection.
The invariant divisor, orthogonal projection and its compatibility with
two independent D3 sectors remain to be specified
(see Appendix~\ref{app:orientifoldscope}). The orbifold-projected charged
representations are then
\begin{align}
\text{E3}\to\text{D3}:&\ (v_0,\bar n_{-a_2/2}), &
\text{D3}\to\text{E3}:&\ (n_{a_2/2},\bar v_0), \notag\\
\text{D7}\to\text{E3}:&\ (u_{a_1/2},\bar v_0), &
\text{E3}\to\text{D7}:&\ (v_0,\bar u_{-a_1/2}).
\label{eq:orbzeromodes}
\end{align}
D3 nodes other than $\pm a_2/2$ and D7 nodes other than $\pm a_1/2$
do not occur in these local charged sectors, and the two nodes in each pair are distinct because $N$
is odd and $a_{1,2}\not\equiv0$. 
We consider a single regular D3-brane and a
single regular D7-brane at the singularity,
with $n_i=u_j=1$ for all $i,j$.
Among the fractional D3 sectors of this
regular representation, we identify the
two sectors relevant to the charged
E3--D3 zero modes as
\begin{equation}
\mathrm{D3}_+\longleftrightarrow
\text{D3 node }-a_2/2,\qquad
\mathrm{D3}_-\longleftrightarrow
\text{D3 node }+a_2/2 .
\label{eq:orbidentification}
\end{equation}
These labels refer here to distinct fractional
D3 sectors at the same orbifold singularity,
rather than to spatially separated D3-branes.
Equation~(\ref{eq:orbzeromodes}) then selects
one Chan--Paton component in each relevant
representation,
\begin{equation}
\begin{aligned}
\lambda_+&=(v_0,\bar n_{-a_2/2}), & \lambda_-&=(n_{a_2/2},\bar v_0),\\
\eta&=(u_{a_1/2},\bar v_0), & \widetilde\eta&=(v_0,\bar u_{-a_1/2}),
\end{aligned}
\label{eq:orbminzeromodes}
\end{equation}
These are the representations needed for Eq.~(\ref{eq:minzeromodes}).
Identifying $\eta$ and $\widetilde\eta$ with one integration variable each
requires one surviving zero mode in each E3--D7 sector. Their charges
restricted to the two selected D3 factors and the diagonal D7 $U(1)$ agree
with Table~\ref{tab:zeromodecharges}. Before recombination the D7-node
$U(1)$s are distinct, and this restricted charge table does not replace the
full quiver gauge group.

With the identification (\ref{eq:orbidentification}) and
$a_1+a_2+a_3=0$, the D3--D7 line of Eq.~(\ref{eq:orbspectra}) gives
\begin{equation}
Q_+=(n_{-a_2/2},\bar u_{a_1/2}),\qquad
\widetilde Q_-=(u_{-a_1/2},\bar n_{a_2/2}).
\label{eq:orbmatter}
\end{equation}
The D7 nodes reached by $Q_+$ and $\widetilde Q_-$ are therefore
automatically those of $\eta$ and $\widetilde\eta$, and the two
Chan--Paton chains
\begin{equation}
\begin{gathered}
E3\xrightarrow{\lambda_+}D3_+\xrightarrow{Q_+}D7_{(a_1/2)}
\xrightarrow{\eta}E3,\\
E3\xrightarrow{\widetilde\eta}D7_{(-a_1/2)}\xrightarrow{\widetilde Q_-}
D3_-\xrightarrow{\lambda_-}E3
\end{gathered}
\label{eq:orbchains}
\end{equation}
close. These are the instanton couplings of type \\
$(\text{E3--D3})$ $(\text{D3--D7})$ $(\text{D7--E3})$ of
Ref.~\cite{IbanezUrangaInstanton}, so the two disks of
Eq.~(\ref{eq:E3disks}) are allowed by the orbifold selection rules. Their
local couplings are of the parent-theory type identified in
Ref.~\cite[Eq.~(2.5)]{IbanezUrangaInstanton}. This establishes the allowed
channel, not the non-vanishing of the complete coefficients $c_\pm$ after
all projections and zero-mode wavefunction overlaps. For the regular Chan--Paton representations chosen here,
the twisted traces are
\[
\mathrm{Tr}\,\gamma_{\theta^k,3}
=\mathrm{Tr}\,\gamma_{\theta^k,7}
=\sum_{j=0}^{N-1}\alpha^{kj}=0,
\qquad k=1,\ldots,N-1.
\]
The local twisted RR tadpole conditions are linear
combinations of these traces, with coefficients
determined by the orbifold twist
\cite[Eq.~(2.18)]{AIQU}.
Since each trace vanishes separately, these conditions
are satisfied for the space-filling D3/D7 content
of this example. This check concerns the local
orbifold background before an orientifold projection.
It does not include orientifold-plane contributions
or global untwisted tadpoles.

\paragraph*{RR charge and physical gauge phase.}
The charge can be evaluated independently of the proposed matter operator
by contracting the D3 couplings to twisted RR two-forms with the E3
couplings to their dual axions. In the conventions of
Refs.~\cite{IbanezRabadanUrangaU1,IbanezUrangaInstanton}, the contraction
contains the oscillator factor $2\sin(\pi k a_2/N)$ and the Chan--Paton
character $n_av_0\alpha^{ka}$ for the chosen E3 node. Using
$2i\sin(\pi k a_2/N)=\alpha^{ka_2/2}-\alpha^{-ka_2/2}$ in that
contraction gives the integer coefficient
\begin{align}
K_a^{(3)}
&=\frac{n_av_0}{N}\sum_{k=0}^{N-1}
\left[\alpha^{k(a+a_2/2)}-\alpha^{k(a-a_2/2)}\right],
\label{eq:RRCharacterSum}\\
\frac{1}{N}\sum_{k=0}^{N-1}\alpha^{km}
&=\delta_{m,0\ ({\rm mod}\ N)}.
\label{eq:orbFourierOrthogonality}
\end{align}
The $k=0$ term in Eq.~(\ref{eq:RRCharacterSum}) vanishes. Including it
only makes the Fourier identity explicit. Consequently,
\begin{equation}
K_a^{(3)}=n_av_0\big(\delta_{0,a+a_2/2}-\delta_{0,a-a_2/2}\big).
\label{eq:IUshift}
\end{equation}
All node indices and Kronecker deltas are understood modulo $N$.
Equations~(\ref{eq:RRCharacterSum})--(\ref{eq:IUshift}) are the
specialization to $v_0=1$ of the RR calculation in
Ref.~\cite[Eqs.~(3.8)--(3.11)]{IbanezUrangaInstanton}, and the coefficient has
not been inferred from the charge of $Q_+\widetilde Q_-$.
In the real gauge-parameter convention of Eq.~(\ref{eq:physicalEshift}),
the corresponding variation of the Euclidean action is purely imaginary:
\begin{equation}
\delta S_E^{\mathrm{cl}}=i\sum_a K_a^{(3)}\alpha_a
\quad\hbox{under D3 transformations}.
\label{eq:orbPhysicalShift}
\end{equation}
With real gauge parameters and the convention
$\Phi\mapsto e^{iq_a\alpha_a}\Phi$, this variation
gives
\[
e^{-S_E^{\mathrm{cl}}}\longmapsto
\exp\!\left[-i\sum_a K_a^{(3)}\alpha_a\right]
e^{-S_E^{\mathrm{cl}}}.
\]
The charge of the instanton weight under the D3-node
factor $U(1)_a$ is therefore $-K_a^{(3)}$.
For $v_0=n_a=1$, the only non-zero coefficients are
$K_{-a_2/2}^{(3)}=+1$ and
$K_{a_2/2}^{(3)}=-1$.
The weight consequently carries charges $(-1,+1)$
under the selected D3 factors, compensating the
charges $(+1,-1)$ of $Q_+\widetilde Q_-$.
Independently, summing the E3--D3 mode charges in
Eq.~(\ref{eq:orbminzeromodes}) gives the same
$(-1,+1)$ charge assignment for the instanton weight.

For a D7-node parameter $\beta_j$, the same character sum with
$n_a\to u_j$, $a\to j$ and $a_2\to a_1$ gives
\begin{align}
K_j^{(7)}&=u_jv_0\big(\delta_{0,j+a_1/2}-\delta_{0,j-a_1/2}\big),\notag\\
\delta S_E^{\mathrm{cl}}\big|_7&=i\sum_jK_j^{(7)}\beta_j,
\label{eq:orbD7Shift}
\end{align}
from Eq.~(3.15) of Ref.~\cite{IbanezUrangaInstanton}. It cancels the
D7-node charges of $Q_+\widetilde Q_-$ as well. The coefficients sum to
zero for $u_j=1$, so the diagonal D7 transformation leaves the weight
invariant. This verifies gauge covariance under the full local node
transformations, without assuming a recombined D7 vacuum.

\paragraph*{Explicit $\mathbb Z_3$ specialization.}
Take $(a_1,a_2,a_3)=(-2,-2,4)$, $\omega=e^{2\pi i/3}$ and
\begin{equation}
\begin{aligned}
\gamma_{\theta,3}=\gamma_{\theta,7}
&=\operatorname{diag}(1,\omega,\omega^2),\\
\gamma_{\theta,E}&=1.
\end{aligned}
\label{eq:Z3CP}
\end{equation}
All node labels below are modulo three. Since $-a_2/2=1$ and $a_2/2=2$,
D3$_+$ is node 1 and D3$_-$ is node 2. Substitution in
Eq.~(\ref{eq:orbspectra}) gives
\begin{align}
\lambda_+&=(v_0,\bar n_1),&\lambda_-&=(n_2,\bar v_0),\notag\\
\eta&=(u_2,\bar v_0),&\widetilde\eta&=(v_0,\bar u_1),\notag\\
Q_+&=(n_1,\bar u_2),&\widetilde Q_-&=(u_1,\bar n_2).
\label{eq:Z3charged}
\end{align}
The two boundary chains are $E\to3_1\to7_2\to E$ and
$E\to7_1\to3_2\to E$. The non-trivial twisted traces vanish because
$1+\omega^k+\omega^{2k}=0$ for $k=1,2$. Equations~(\ref{eq:IUshift})
and~(\ref{eq:orbD7Shift}) give
\begin{equation}
\delta S_E^{\mathrm{cl}}
=i(\alpha_1-\alpha_2+\beta_1-\beta_2).
\label{eq:Z3shift}
\end{equation}
The matter product transforms by the opposite phase to $e^{-S_E^{\mathrm{cl}}}$,
namely $e^{i(\alpha_1-\alpha_2+\beta_1-\beta_2)}$.
The diagonal D7 choice $\beta_0=\beta_1=\beta_2$ reduces this to
Eq.~(\ref{eq:physicalEshift}). The local E3--D7 components in
Eq.~(\ref{eq:Z3charged}) are still subject to the zero-mode condition above.

\paragraph*{Additional matter and saturation channels.}
The regular representations also contain D3 node 0 and
additional matter multiplets. In particular, the fields
with the D3 charges of the prospective Dirac partners are
$\widetilde Q_+=(u_0,\bar n_1)$ and $Q_-=(n_2,\bar u_0)$,
as follows from the third line of Eq.~(\ref{eq:orbspectra}).
Identifying these representations does not specify Dirac
mass terms. The calculation retains the local quiver
spectrum and selects the charged representations entering
the two disk channels, without assuming a truncation to
two massive fermions.

There is also a coupling $\lambda_+\Phi^{(2)}\lambda_-$ to the D3--D3
bifundamental $\Phi^{(2)}_{-a_2/2,\,a_2/2}$ associated with $z_2$
\cite[Eq.~(2.3)]{IbanezUrangaInstanton}. It absorbs both $\lambda$
variables and therefore provides an alternative to the two insertions
generating $Q_+\widetilde Q_-$. If the relevant D7 chiral mode survives,
a coupling of type $(\text{E3--D7})(\text{D7--D7})(\text{D7--E3})$
can absorb $\eta,\widetilde\eta$ as well
\cite[Eq.~(2.6)]{IbanezUrangaInstanton}. Saturating these two couplings
then produces another matter bilinear, involving a D3--D3 field and a
D7--D7 field. This possible channel does not remove the two-disk term,
it illustrates why the charged integral alone is not a truncation of the
four-dimensional quiver to two massive fermions.

\subsection{Orientifold constraint and scope of the example}

\label{app:orientifoldscope}

\paragraph*{Orientifold constraint.}
The orbifold projection used above does not by itself specify
an orientifold projection. For the conventional orientifold
action conjugating Chan--Paton phases, node $i$ maps to
node $-i$ \cite{IbanezUrangaInstanton}. At a fixed singularity,
this action identifies the selected nodes $\pm a_2/2$,
so they cannot represent the two independent D3 gauge
sectors assumed in the operator calculation.

Ref.~\cite[Sec.~2.1]{IbanezUrangaInstanton} discusses a
different setting in which a singularity and its distinct
orientifold image are intersected by an invariant E3.
This illustrates why the orientifold action on the geometry,
as well as on the Chan--Paton labels, matters for the
surviving sectors. The choice $v_0=1$ in the present example
specifies the orbifold Chan--Paton representation but does
not establish the neutral $O(1)$ measure or its compatibility
with two independent D3 sectors. Likewise, the point reflection
$z_i\to-z_i$ exchanges the two positions in
Eq.~(\ref{eq:branePositions}) and does not retain them as
independent sectors in the intended sense.

\paragraph*{Matter content and endpoint geometry.}
The selected fractional D3 sectors lie at the same orbifold
singularity, through which the D7 passes. The resulting
massless quiver spectrum includes additional gauge factors
and matter multiplets, as described above. It is therefore
distinct from the massive two-Dirac-fermion sector used
in the main text. The same distinction applies to the endpoint
geometry. Restricted to the diagonal D7 gauge
field, the coincident endpoints and trivial path
give $\mathcal A_\mu=0$ in
Eq.~(\ref{eq:relativeConnection}).
The full quiver instead contains distinct D7-node
gauge factors, while the matching in
Sec.~\ref{sec:relative} requires separated endpoints.

\paragraph*{Scope of the illustrative calculation.}
The example explicitly determines the local charged
representations and allowed Chan--Paton disk channels,
checks local orbifold twisted tadpole cancellation, and
derives the compensating RR charge. It thus illustrates
the relation between a chiral E3--D3 spectrum and the
gauge transformation of the instanton weight. These
results provide a complementary check of the charged
ingredients entering the proposed mass operator.

The distinction between representations and integration
variables remains essential for interpreting this check.
The E3--D7 representations must supply the assumed
zero-mode multiplicities, and the full instanton
contribution also requires the neutral
$d^4x\,d^2\theta$ measure and non-vanishing charged disk
amplitudes \cite{IbanezUrangaInstanton,BCKW}.
These are inputs to the conditional saturation calculation.
They are not established by the orbifold projection and
RR-charge matching alone.

\section{Finite-separation BCFT factor}

\label{app:finiteBCFT}

This appendix isolates the world-sheet free-boson sector directly sensitive
to the change of the D3--D7 Dirichlet separation in the real $x^8$ direction
of $z_3$. It does not evaluate the full charged superstring disk amplitude of Eq.~(%
\ref{eq:E3disks}). Its purpose is to test whether this particular
finite-separation factor introduces a new zero. The general use of DN
boundary twist fields in open-string amplitudes is standard \cite%
{HashimotoDN} and we use the exact two-twist/one-primary correlator reviewed
in Ref.~\cite{MattielloSachs}.

For either charged disk the $x^8$ boundary is divided into three intervals.
In cyclic order their boundary conditions are N on the E3, D at $x^8=a_s$ on
D3$_s$, and D at $x^8=0$ on the D7, where 
\begin{equation}
a_+=-\ell,\qquad a_-=+\ell.  \label{eq:asBCFT}
\end{equation}
The discontinuity between the two Dirichlet values is precisely the
classical stretch of the $37$ state. Compactify $x^8$ temporarily on a
circle much larger than all local scales and T-dualize it. T-duality is an
exact BCFT equivalence and exchanges N and D
\cite{Polchinski:1998rq,Bachas:1998rg}. The two D intervals become
Neumann intervals carrying different Wilson lines, so their
boundary-changing open string is represented in this real-boson sector by an
ordinary momentum primary with 
\begin{equation}
k_s=\epsilon_s\frac{\ell}{2\pi\alpha^{\prime}},\qquad \epsilon_s=\pm1,
\label{eq:ksBCFT}
\end{equation}
where the sign is fixed by the orientation of the corresponding $37$ or $73$
string. The E3--D3 and E3--D7 changes become a conjugate pair of DN/ND
twists. The large-radius limit can then be taken after the local correlator
is evaluated.

Let $X$ denote the canonically normalized chiral boson used in Ref.~\cite%
{MattielloSachs}, with $\psi_\alpha\propto:e^{i\alpha X}:$ and $%
h_\alpha=\alpha^2/2$. Matching to the open-string momentum contribution
gives 
\begin{equation}
h_{\alpha_s}=\frac{\alpha_s^2}{2}=\alpha^{\prime}k_s^2 =\frac{\ell^2}{%
4\pi^2\alpha^{\prime}}.  \label{eq:halphaBCFT}
\end{equation}
For a fixed T-dual Dirichlet/Wilson-line sector, the exact boundary
correlator in the presence of one Dirichlet sector is 
\begin{equation}
\left\langle\bar\sigma(x_1)\psi_{\alpha_s}(x_2) \sigma(x_3)\right\rangle
=e^{i\alpha_s x_0} \frac{x_{31}^{\,h_{\alpha_s}-1/8}} {x_{21}^{\,h_{%
\alpha_s}}x_{32}^{\,h_{\alpha_s}}},  \label{eq:BCFTappendix3pt}
\end{equation}
where $x_0$ is the Dirichlet zero mode of the fixed twist sector. This
formula is just the conformal three-point function with the non-zero
structure constant $e^{i\alpha_s x_0}$. In particular, no zero-mode momentum
delta function is present in that fixed DN sector. For ordered, distinct
boundary insertion points the correlator is therefore non-zero for every
finite $\alpha_s$. Changing $k_s\to-k_s$ conjugates its zero-mode phase but
cannot make it vanish. This statement does not replace the compact
neutral-zero-mode analysis: if the corresponding continuous Wilson-line
datum survives as a dynamical instanton modulus, its integration may impose
an additional selection rule, so the intended minimal $O(1)$ sector must
project out or lift such a modulus.

The $x^9$ coordinate has no relative displacement. Chan--Paton closure and
total picture number were checked in Eqs.~(\ref{eq:E3CPclosure}) and 
in Appendix~\ref{app:CFT}. The result of the present appendix is therefore the
narrow statement encoded in Eq.~(\ref{eq:AfiniteNonzero}): the isolated
separated-$x^8$ factor $F_s^{(8)}(\ell)$ is non-zero at every finite
$\ell$. The remaining factors entering the complete disk coefficient $c_s$
are not proven non-zero here. Consequently no vanishing can be attributed to
this free-boson DD-separation factor alone, while the full charged-disk
condition Eq.~(\ref{eq:fullDiskCondition}) is an input of the conditional
mechanism, to be tested in a specified realization.

\section{From the Weyl off-diagonal mass term to the TBTS Dirac matrix}

\label{app:Diracconversion}

Let $\chi_s$ and $\widetilde\chi_s$ be left-handed Weyl fermions. 
Using the conventions introduced in Eq.~(\ref{eq:D3D7Dirac}), we write in the Weyl basis
\begin{equation}
\Psi_s=
\begin{pmatrix}
\chi_s \\
\widetilde\chi_s^\dagger
\end{pmatrix},
\qquad
\bar\Psi_s=
\begin{pmatrix}
\widetilde\chi_s & \chi_s^\dagger
\end{pmatrix}.
\label{eq:Diracappendixdef}
\end{equation}
With $P_{L,R}=(1\mp\gamma^5)/2$, one then has
\begin{equation}
\chi_s\widetilde\chi_t
=
\bar\Psi_t P_L\Psi_s,
\qquad
\left(\chi_s\widetilde\chi_t\right)^\dagger
=
\bar\Psi_s P_R\Psi_t.
\label{eq:WeylDiracidentity}
\end{equation}
Consequently, the single instanton-induced off-diagonal mass term gives
\begin{equation}
-\mu_{+-}\chi_+\widetilde\chi_-+\mathrm{h.c.}
=
-\mu_{+-}\bar\Psi_-P_L\Psi_+
+\mathrm{h.c.}.
\label{eq:crossDiracappendix}
\end{equation}
The coefficient may subsequently be made real and positive by the same
residual vector rephasing used in Sec.~\ref{sec:gamma5}.

For
\begin{equation}
\Psi_-^{\mathrm{TBTS}}
=
i\gamma^5\Psi_-^{\mathrm{str}},
\label{eq:appendixaxial}
\end{equation}
the Dirac adjoint transforms as
\begin{equation}
\bar\Psi_-^{\mathrm{TBTS}}
=
i\bar\Psi_-^{\mathrm{str}}\gamma^5,
\qquad
\bar\Psi_-^{\mathrm{TBTS}}
\Psi_-^{\mathrm{TBTS}}
=
-\bar\Psi_-^{\mathrm{str}}
\Psi_-^{\mathrm{str}}.
\label{eq:appendixbars}
\end{equation}
Thus the axial redefinition reverses the sign of the scalar mass
bilinear, while the vector current remains invariant.

Solving Eq.~(\ref{eq:appendixaxial}) for the string-basis field and
using $\gamma^5P_L=-P_L$ gives
\begin{equation}
-\mu_{+-}\bar\Psi_-^{\mathrm{str}}P_L\Psi_+
=
\frac{i\mu_{+-}}{2}
\bar\Psi_-^{\mathrm{TBTS}}
(\gamma^5-1)\Psi_+.
\label{eq:appendixgamma5}
\end{equation}
Comparing this expression with the TBTS off-diagonal interaction,
Eq.~(\ref{eq:LTBTSmix}), immediately yields
\begin{equation}
g=\frac{\mu_{+-}}{2},
\qquad
m_r=-\frac{\mu_{+-}}{2}=-g,
\end{equation}
in agreement with Eq.~(\ref{eq:gmfrommu}).

\section{Soft D7-vector dressing of the chiral mass vertex}

\label{app:softVectorDressing}

This appendix examines the response of the scalar form
factor $G_E[A]$ to a soft four-dimensional D7 vector
fluctuation. The aim is to determine whether this
additional dressing changes the leading
$O(k^0\delta A)$ contribution of the Wilson-completed
mass operator used in Sec.~\ref{sec:chiralDressing}.
The argument combines four-dimensional Lorentz
invariance with a local expansion in the soft momentum,
while keeping the internal gauge-mode profile fixed.

Let us choose a reference background $A^{(0)}$ with $A_\mu^{(0)}=0$ and all internal
background fields independent of $x^\mu$. We also choose the reference gauge
along $\gamma$ so that $U_\gamma[A^{(0)}]=1$. Any static path phase is thereby
absorbed into the definition of the reference mass. The reference instanton
contribution is already contained in the canonically normalized mass
$\mu_{+-}$. We collect all additional gauge-field dependence of this chiral
mass term into a dimensionless, gauge-invariant form factor $G_E[A]$,
normalized by $G_E[A^{(0)}]=1$. The corresponding Wilson-completed
off-diagonal block is already displayed in
Eq.~(\ref{eq:Lstringmassfunctional}). Here we only analyze the additional
soft-field dependence carried by $G_E[A]$.

We work with a local Wilsonian vertex after integrating out heavy string,
KK and intersection excitations, while retaining the D7 vector modes used
as external fields. The response of its coefficients is analytic in soft
four-dimensional momentum below the integrated-out thresholds
\cite{BurgessEFT}. The reference background preserves four-dimensional
Lorentz invariance. Since $G_E$ is a Lorentz scalar with no derivatives
acting on the external matter fields, a term linear in a four-dimensional
vector fluctuation requires at least one four-dimensional derivative:
internal background tensors alone cannot contract its Lorentz index.
Gauge invariance further restricts the allowed terms. Denoting the soft four-momentum scale by $k$, we expand
to first order in the vector fluctuation,
\begin{equation}
G_E[A^{(0)}+\delta A]
=1+\delta G_E^{(1)}+O((\delta A)^2).
\label{eq:Wilsontruncation}
\end{equation}
The Lorentz-index argument above then gives
\begin{equation}
\delta G_E^{(1)}=O(k\,\delta A),\qquad
\partial_\mu\delta G_E^{(1)}=O(k^2\delta A).
\label{eq:Gpowercount}
\end{equation}
These estimates are power-counting bounds. Gauge
invariance may impose further suppression or make
the linear response vanish.
This counting concerns the scalar coefficient of the
mass operator, with no derivatives acting on the
external matter fields. Nonlocal contributions from
the propagation of retained light fields are not
included in this Wilsonian coefficient
\cite{BurgessEFT}.

For completeness, consider one four-dimensional D7 vector mode,
\begin{align}
A_M&=A_M^{(0)}+\delta A_M,  \notag \\
\delta A_\mu(x,y)&=f_n(y)a_\mu(x),\qquad \delta A_a=0,
\label{eq:softvectormode}
\end{align}
with four-momentum $k_\mu$. Its linearized field strengths are
\begin{align}
\delta F_{\mu a}&=-\partial_a f_n\,a_\mu,  \notag \\
\delta F_{\mu\nu}&=f_n(\partial_\mu a_\nu-\partial_\nu a_\mu),\qquad
\delta F_{ab}=0.
\label{eq:softvectorF}
\end{align}
At fixed internal profile $f_n$, the mixed component
$\delta F_{\mu a}$ contains no four-dimensional
derivative and can remain non-zero as $k\to0$.
The suppression of the linear response of $G_E$
therefore follows from its scalar Lorentz structure,
not from counting every field strength as a power
of $k$. An operator containing a fermion current
has additional Lorentz indices available for
contraction and is not subject to the same argument.
The result established here consequently concerns
the scalar dressing of the mass operator, rather
than the complete amplitude with a D7 vector insertion.
Equation~(\ref{eq:relativeConnection}) gives
\begin{equation}
\delta \mathcal{A}_\mu=
\left[f_n(p_+)-f_n(p_-)\right]a_\mu(x),
\label{eq:pathindependentvector}
\end{equation}
which is independent of the chosen reference path at linear order in this
vector sector. This is the mode-by-mode form of the relative connection in
Eq.~(\ref{eq:relativeModeExpansion}).

It is also unnecessary to assume that $G_E[A]$ is real away from the reference
background. Define
\begin{equation}
\mu_E[A]\equiv\mu_{+-}G_E[A],\qquad
\mu_E[A^{(0)}]=\mu_{+-}.
\label{eq:gmfunctional}
\end{equation}
Applying the axial field redefinition~(\ref{eq:axialredef}) directly to
Eq.~(\ref{eq:Lstringmassfunctional}) gives
\begin{equation}
\mathcal{L}_{\mathrm{off}}^{\mathrm{TBTS}}[A]=
\frac{i}{2}\mu_E[A]\,\bar\Psi_-U_\gamma(\gamma^5-1)\Psi_+
+\mathrm{h.c.}
\label{eq:LTBTSmixfunctional}
\end{equation}
At the reference background, the residual vector rephasing makes $\mu_{+-}$
real and positive and Eq.~(\ref{eq:LTBTSmixfunctional}) reduces to
Eqs.~(\ref{eq:LTBTSmix}) and~(\ref{eq:gmfrommu}), with $m_r=-g$. For a soft
vector fluctuation,
\begin{equation}
\delta\mu_E^{(1)}=\mu_{+-}\,\delta G_E^{(1)}=O(k\delta A),\qquad
\partial_\mu\delta\mu_E^{(1)}=O(k^2\delta A).
\label{eq:gmfunctionalpower}
\end{equation}
Thus possible changes of the scalar/pseudoscalar decomposition begin only in
the subleading form-factor response.

Before taking the leading soft limit, define
$J_5^\mu[A]=\bar\Psi_f\gamma^\mu\gamma^5U_\gamma G_E\Psi_i$ and
$P_5[A]=\bar\Psi_f\gamma^5U_\gamma G_E\Psi_i$. The Dirac equations give
\begin{align}
\partial_\mu J_5^\mu[A]={}&2imP_5[A]
+iq\bar\Psi_f\gamma^5\slashed{\mathcal{A}}U_\gamma G_E\Psi_i  \notag \\
&+\bar\Psi_f\gamma^\mu\gamma^5U_\gamma(\partial_\mu G_E)\Psi_i.
\label{eq:stringaxialfull}
\end{align}
By Eq.~(\ref{eq:Gpowercount}), the last term begins at $O(k^2\delta A)$ in
the vector sector. The field-dependent correction to the chiral coefficient
itself starts at $O(k\delta A)$. Consequently neither contribution modifies
the $O(k^0\delta A)$ contribution of the mass operator kept in
Eqs.~(\ref{eq:stringaxial})--(\ref{eq:leadingWilsonAmplitude}).

\section{Signs in the Wilson-line axial identity}

\label{app:identity}

This appendix checks the signs of the pseudoscalar Wilson identity at the
leading soft-vector order~(\ref{eq:Wilsontruncation}). Before taking that
order, the additional derivative term proportional to $\partial_\mu G_E$ is
displayed explicitly in Eq.~(\ref{eq:stringaxialfull}). By
Eq.~(\ref{eq:Gpowercount}) it begins at $O(k^2\delta A)$ for the vector
sector considered here. The scalar partner $m_r=-g$ is a distinct transition
channel and is not part of this sign check.

For completeness, take $D_\mu=\partial_\mu+iqA_\mu$ and $\Psi\to
e^{-iq\alpha}\Psi$. The endpoint equations imply
\begin{align}
\gamma^\mu\partial_\mu\Psi_i&=-i(m+q\slashed A_+)\Psi_i,  \notag \\
(\partial_\mu\bar\Psi_f)\gamma^\mu&=i\bar\Psi_f(m+q\slashed A_-).
\label{eq:appEOM}
\end{align}
Differentiating $J_5^\mu=\bar\Psi_f\gamma^\mu\gamma^5U_\gamma\Psi_i$
produces the external-leg term
\begin{equation}
2imP_5+iq\bar\Psi_f\gamma^5(\slashed A_+-\slashed A_-)U_\gamma\Psi_i,
\label{eq:appLegs}
\end{equation}
and the derivative of the Wilson transporter. Since the D7 field is Abelian,
\begin{equation}
\frac{1}{iq}U_\gamma^{-1}\partial_\mu U_\gamma
=\int_\gamma \partial_\mu A_a\,dy^a.
\label{eq:appWilsonTransport}
\end{equation}
Combining this with the external-leg term gives Eq.~(\ref{eq:stringaxial})
with
\begin{align}
\mathcal{A}_\mu(\gamma)
&=A_\mu(p_+)-A_\mu(p_-)-\frac{1}{iq}U_\gamma^{-1}\partial_\mu U_\gamma
\notag \\
&=\int_\gamma(\partial_aA_\mu-\partial_\mu A_a)\,dy^a
=-\int_\gamma F_{\mu a}\,dy^a,
\label{eq:appStokes}
\end{align}
fixing the sign in Eq.~(\ref{eq:relativeConnection}).

\end{document}